\documentclass[aps,prl,preprintnumbers,showpacs,nofootinbib,linenumbers]{revtex4}
\usepackage{booktabs}
\usepackage{siunitx}
\usepackage[utf8]{inputenc}
\usepackage[english]{babel}
\usepackage{amsmath}
\usepackage{amsfonts}
\usepackage{amssymb}
\usepackage{epsfig}
\usepackage{graphics,psfrag,rotating}
\usepackage{graphicx}
\usepackage{dcolumn}
\usepackage{bm}
\usepackage{epstopdf}
\usepackage{color}
\usepackage{float}
\usepackage[usenames,dvipsnames,svgnames]{xcolor}
\usepackage[colorlinks=true,
            linkcolor=red,
            urlcolor=gray,
            citecolor=blue]{hyperref}
  \usepackage{hyperref}
    \usepackage{lineno}

\newcommand{\lsim}   {\mathrel{\mathop{\kern 0pt \rlap
{\raise.2ex\hbox{$<$}}}
 \lower.9ex\hbox{\kern-.190em $\sim$}}}
\newcommand{\gsim}   {\mathrel{\mathop{\kern 0pt \rlap
{\raise.2ex\hbox{$>$}}}
\lower.9ex\hbox{\kern-.190em $\sim$}}}

\def\3nab{\tilde{\nabla}}

\def\hsp5{\hspace{5mm}}

\def\case#1/#2{\textstyle\frac{#1}{#2}}

\def\ber {\begin{eqnarray}}
\def\eer {\end{eqnarray}}
\def\bea {\begin{eqnarray}}
\def\eea {\end{eqnarray}}

\def\bc {\begin{center}}
\def\ec {\end{center}}
\def\case#1/#2{\frac{#1}{#2}}

\newcommand{\bw}{\begin{widetext}}
\newcommand{\ew}{\end{widetext}}

\newcommand{\be}{\begin{equation}}
\newcommand{\bse}{\begin{subequation}}
\newcommand{\ese}{\end{subequation}}
\newcommand{\ee}{\end{equation}}
\newcommand{\eei}{\end{eqnarray}\indent\indent}
\newcommand{\ba}{\begin{array}}
\newcommand{\ea}{\end{array}}
\newcommand{\bal}{\begin{eqnarray}}
\newcommand{\eal}{\end{eqnarray}}

\def\case#1/#2{\textstyle\frac{#1}{#2} }

\newcommand{\bver}{\begin{verbatim}}
\newcommand{\ever}{\end{verbatim}}

\begin{document}


\title{ DESI DR2 Constraints on Coupled Hu--Sawicki $f(R)$ Gravity with Interacting Neutrinos: Implications for the Hubble and  \(S_8\) Tensions}
\author{
 Muhammad Yarahmadi$^{1}$\footnote{Email: yarahmadimohammad10@gmail.com, yarahmadi.mhd@lu.ac.ir}
}
\affiliation{Department of Physics, Lorestan University, Khoramabad, Iran}

\date{\today}

\begin{abstract}
	We investigate the cosmological implications of viable Hu--Sawicki \(f(R)\) gravity models and their extension through an effective interaction with the neutrino sector. Motivated by the persistent \(H_0\) and \(S_8\) tensions, we explore whether the combined effects of modified gravity, neutrino free-streaming, and dark-sector interaction can alleviate these discrepancies while remaining consistent with current cosmological observations. Starting from the metric formulation of \(f(R)\) gravity, we derive the modified background and perturbation equations and implement the resulting cosmological scenario within the {MontePython}/{hi\_class} framework. A comprehensive Bayesian MCMC analysis is performed using combinations of Planck 2018 CMB data, CMB lensing, DESI DR2 BAO measurements, Cosmic Chronometers, and Pantheon supernova observations. Our results show that the interacting Hu--Sawicki scenario provides a noticeable improvement over both the standard \(\Lambda\)CDM cosmology and the non-interacting \(f(R)\) framework. In particular, the interacting model shifts the inferred Hubble constant toward larger values, yielding \(H_0 = 70.21 \pm 1.60~{\rm km~s^{-1}Mpc^{-1}}\), while simultaneously lowering the clustering amplitude to \(S_8 = 0.796 \pm 0.009\). Consequently, the residual tensions with SH0ES R22 and DES-Y6 are reduced to nearly the \(1\sigma\) level. From the statistical perspective, the interacting model provides the best overall fit according to the Akaike Information Criterion, with \(\Delta\chi^2 \simeq -45\) relative to \(\Lambda\)CDM. The obtained neutrino mass constraints remain fully consistent with current cosmological bounds, with \(\sum m_\nu < 0.122~{\rm eV}\). These results suggest that interacting modified-gravity scenarios offer a viable route to moderately easing both late-time tensions at once.
\end{abstract}


%
%


\maketitle

\section{Introduction}
\label{sec:intro}

The determination of the present cosmic expansion rate, quantified by the Hubble constant \(H_0\), is one of the central challenges of precision cosmology. Over the past decade, increasingly accurate observations have revealed a persistent and statistically significant discrepancy between the values of \(H_0\) inferred from early-Universe probes and those measured directly in the late Universe. This discrepancy, commonly referred to as the \emph{Hubble tension}, is now widely regarded as a potential indication of physics beyond the standard \(\Lambda\)CDM paradigm rather than a mere consequence of unresolved observational systematics. The most precise early-Universe determination of \(H_0\) arises from measurements of the cosmic microwave background (CMB) anisotropies. Using observations of the acoustic peak structure in the temperature and polarization power spectra, the \textit{Planck} Collaboration reported $H_0 = 67.4 \pm 0.5~\mathrm{km\,s^{-1}\,Mpc^{-1}}$, within the framework of the spatially flat \(\Lambda\)CDM model~\cite{Planck18}. This cosmological scenario also provides a consistent description of a broad range of early-Universe observables, including primordial nucleosynthesis, baryon acoustic oscillations, and large-scale structure formation. Nevertheless, direct low-redshift measurements consistently favor substantially larger values of \(H_0\). A leading local determination has been obtained by the SH0ES (Supernovae \(H_0\) for the Equation of State) collaboration through a refined cosmic distance ladder based on Cepheid-calibrated Type Ia supernovae observed with the Hubble Space Telescope. Their latest analysis yields $H_0 = 73.04 \pm 1.04~\mathrm{km\,s^{-1}\,Mpc^{-1}}$, corresponding to a discrepancy exceeding \(4.9\sigma\) relative to the \textit{Planck} result~\cite{Riess2022}. Independent late-Universe probes reinforce this tension. The Carnegie--Chicago Hubble Program (CCHP), employing the Tip of the Red Giant Branch (TRGB) calibration, obtained $H_0 = 69.6 \pm 0.8 \pm 1.7~\mathrm{km\,s^{-1}\,Mpc^{-1}}$, while strong-lensing time-delay analyses by the H0LiCOW collaboration reported $H_0 = 73.3^{+1.7}_{-1.8}
\quad \text{and} \quad 75.3^{+3.0}_{-2.9}~\mathrm{km\,s^{-1}\,Mpc^{-1}}$ from independent lens systems~\cite{FreedmanTRGB20,Wonglens19,Weilens20}. Additional determinations based on CMB lensing and Mira variable stars provide $H_0 = 73.5 \pm 5.3~\mathrm{km\,s^{-1}\,Mpc^{-1}}$ and $H_0 = 72.7 \pm 4.6~\mathrm{km\,s^{-1}\,Mpc^{-1}}$, respectively~\cite{BaxterCMBlens20,HuangMiras19}. The consistency among several independent late-time probes suggests that the Hubble tension reflects a genuine physical inconsistency between early- and late-Universe cosmology, and not merely a calibration artifact~\cite{VerdeTR19}. Comprehensive reviews of cosmological tensions and their possible theoretical interpretations can be found in~\cite{DiValentino2021,TensionsReview2025}. A related discrepancy has also emerged in measurements of the amplitude of matter clustering, commonly quantified through the parameter $S_8 \equiv \sigma_8 \sqrt{\Omega_m/0.3}$. While CMB observations inferred from \textit{Planck} favor relatively large clustering amplitudes, weak-lensing and galaxy-survey measurements have historically indicated systematically smaller values. The current status of this so-called \emph{\(S_8\) tension}, however, has evolved considerably with the latest observational analyses. In particular, recent studies by the KiDS collaboration suggest that the previously reported discrepancy with \textit{Planck} has been substantially reduced and may now be compatible within the \(1\sigma\) level~\cite{KiDS2025A,KiDS2025B}. Likewise, the latest DES-Y6 weak-lensing analysis provides significantly improved cosmological constraints relative to earlier DES-Y3 results~\cite{DESY6_2026}. Although the statistical significance of the \(S_8\) discrepancy remains under active investigation, these observations continue to motivate extensions of the standard cosmological framework, especially in scenarios involving modified gravity, interacting dark sectors, or nonstandard neutrino physics. Together with the persistent \(H_0\) discrepancy, these tensions may provide important clues regarding the fundamental physics governing the late-time Universe. A wide variety of theoretical approaches have been proposed to alleviate these inconsistencies. One possible direction involves extending the dark-energy sector through dynamical scalar-field models, including quintessence~\cite{agr-cosmo-de-quintessence-01}, tachyonic fields~\cite{agr-cosmo-de-tachyon-01}, k-essence~\cite{agr-cosmo-de-k-essence-01}, phantom energy~\cite{agr-cosmo-de-phantom-01}, Chaplygin gas~\cite{agr-cosmo-de-chaplygin-01}, as well as holographic and agegraphic dark-energy scenarios~\cite{agr-cosmo-de-holo-01,agr-cosmo-de-holo-02,agr-cosmo-de-holo-04,agr-cosmo-de-agegraphic-01,agr-cosmo-de-agegraphic-02}. A further possibility is to modify the geometric sector of General Relativity (GR) itself. In this context, numerous modified gravity theories have been investigated, including \(f(R)\) gravity~\cite{agr-modified-gravity-fR-review-01}, \(f(T)\) gravity~\cite{agr-modified-gravity-fT-review-01}, \(f(R,T)\) gravity~\cite{agr-modified-gravity-fRT-01}, Brans--Dicke theory~\cite{agr-modified-gravity-BD-01}, Gauss--Bonnet and Lovelock gravities~\cite{agr-modified-gravity-GB-01,agr-modified-gravity-Lovelock-01}, and Ho\v{r}ava--Lifshitz gravity~\cite{agr-modified-gravity-horava-01,agr-modified-gravity-horava-02,agr-modified-gravity-horava-lw-01,agr-modified-gravity-horava-lw-05}. Modified gravity scenarios have been extensively explored as possible mechanisms for alleviating the Hubble tension by altering either the cosmic expansion history or the growth of structure~\cite{Frusciante2020,Basilakos2022}. In particular, scalar--tensor models formulated in the Jordan frame with nonminimal couplings between scalar fields and the metric have attracted considerable attention in late-time cosmology~\cite{JordanFrame2022,JordanFrame2021,JordanFrame2022b}. Among the various modified gravity candidates, \(f(R)\) gravity is widely studied for its conceptual simplicity and phenomenological richness. In these models, the Ricci scalar \(R\) appearing in the Einstein--Hilbert action is generalized to an arbitrary function \(f(R)\), introducing an additional scalar degree of freedom that can effectively mimic dark energy~\cite{Nojiri1,Nojiri2,Sotiriou2010,De Felice,Capozziello1,Capozziello2,Capozziello3}. Within the metric formalism, \(f(R)\) gravity preserves diffeomorphism invariance while naturally generating scale-dependent modifications to the gravitational interaction, leading to distinctive signatures in both the background expansion and cosmological perturbations. Further discussions on the theoretical foundations and cosmological implications of modified gravity, including \( f(R) \) models, can be found in \cite{ Sotiriou2010, Nojiri2011}.
Among the viable modifications of General Relativity, the Hu--Sawicki $f(R)$ gravity model~\cite{HuSawicki2007} provides a widely used framework for explaining late-time cosmic acceleration while evading local gravity constraints. By construction, this phenomenological model asymptotically recovers standard Einsteinian gravity in the high-curvature limit, restricting dynamical deviations exclusively to cosmological scales. The functional form of the model is expressed as
\begin{equation}
	f(R) = R - m^2 \frac{c_1 (R/m^2)^n}{c_2 (R/m^2)^n + 1},
\end{equation}
where $m^2$ establishes the characteristic curvature scale, while $c_1$, $c_2$, and $n$ are free parameters that collectively dictate the magnitude and steepness of the deviation from General Relativity. In the high-curvature regime (\(R \gg m^2\)), the model approaches $f(R)\simeq R-2\Lambda_{\mathrm{eff}}$, thereby recovering the \(\Lambda\)CDM limit and satisfying local gravity constraints. At low curvatures, however, nonlinear corrections become important and can naturally drive the late-time accelerated expansion without the need for an explicit cosmological constant. This flexibility makes the HS model a useful phenomenological framework for exploring late-Universe tensions.

A further extension of this scenario arises when the modified gravitational sector is coupled to massive neutrinos. Massive neutrinos play a fundamental role in cosmology through their impact on the expansion history, the anisotropy evolution of the CMB, and the suppression of matter clustering on small scales~\cite{Lesgourgues2014,NeutrinoCosmo2024}. Owing to their weak interactions and relativistic-to-nonrelativistic transition during cosmic evolution, neutrinos are sensitive probes of deviations from standard gravitational dynamics. An interaction between the curvature scalar degree of freedom, $f_R \equiv \frac{df}{dR}$, and the neutrino sector can modify both the background cosmological evolution and the behavior of perturbations. Such a coupling effectively allows energy exchange between the curvature and neutrino components, which in turn affects the sound horizon, structure growth, and clustering amplitude. The interplay between modified gravity and massive neutrinos is of particular interest because neutrino free-streaming tends to suppress matter clustering, partially compensating for the enhanced growth typically induced by \(f(R)\) gravity~\cite{fRNu2025}. Motivated by this complementary behavior, we investigate an effective interaction between the neutrino and gravitational sectors in order to explore whether such a framework can simultaneously alleviate both the \(H_0\) and \(S_8\) tensions. In this work, we study the cosmological implications of the Hu--Sawicki model of \(f(R)\) gravity in the presence of a curvature--neutrino interaction. We investigate how this coupling modifies the late-time expansion history and the evolution of cosmological perturbations, and we assess its viability through a detailed comparison with current observational datasets. Our analysis combines analytical derivations, dynamical-system techniques, and Bayesian statistical constraints to evaluate the capability of the model to reconcile the observed discrepancies between early- and late-Universe cosmological measurements.
\section{$f(R)$ gravity model}
$f(R)$ gravity is a theoretically well-motivated extension of Einstein’s General Relativity that modifies the Einstein–Hilbert action by replacing the Ricci scalar $R$ with a general function $f(R)$. This modification introduces an additional scalar degree of freedom—often referred to as the “scalaron”—which alters the effective gravitational dynamics and enables a purely geometric explanation of the Universe’s late-time acceleration \cite{De Felice, Bergmann, Liu, Buchdahl}. Unlike the cosmological constant, the curvature-dependent modifications in $f(R)$ gravity do not generically act as a repulsive force; rather, their dynamical behavior strongly depends on the specific functional form of $f(R)$. In viable models, such as the Hu--Sawicki class, the functional form is constructed so that deviations from General Relativity are strongly suppressed in the high-curvature regime (corresponding to early times and high redshifts), while becoming relevant only at low curvature scales, thereby driving late-time cosmic acceleration.

Moreover, consistency with Solar-System tests is not automatic and requires the presence of a screening mechanism. In particular, viable $f(R)$ models satisfy local gravity constraints through the Chameleon mechanism, in which the scalar degree of freedom (scalaron) acquires an environment-dependent effective mass. In high-density regions such as the Solar System, the scalaron becomes sufficiently massive, suppressing fifth-force effects and effectively restoring General Relativity. The viability of a given model therefore depends on ensuring that the scalaron mass satisfies local constraints for the specific functional form of $f(R)$.

From a cosmological perspective, $f(R)$ gravity offers a unified framework capable of addressing both the observed accelerated expansion and current observational tensions such as those in $H_0$ and $S_8$. The modified field equations affect the evolution of density perturbations and the growth of large-scale structures, yielding distinct signatures in weak-lensing, BAO, and CMB observations. Moreover, by introducing a scalar mode with scale-dependent dynamics, $f(R)$ gravity provides a natural bridge between modified gravity theories and the effective field theory of dark energy.

The action for $f(R)$ gravity in the presence of matter components is given by
\begin{equation}
	S = \int d^4x \sqrt{-g} \left( \frac{f(R)}{16\pi G} + \mathcal{L}_m \right),
\end{equation}
where $R$ is the Ricci scalar and $\mathcal{L}_{m}$ denotes the matter Lagrangian.

Varying the action with respect to the metric tensor $g_{\mu\nu}$ yields the field equations:
\begin{equation}
	f_{R} R_{\mu\nu} - \frac{1}{2} f(R) g_{\mu\nu}
	- \nabla_{\mu}\nabla_{\nu} f_{R}
	+ g_{\mu\nu} \Box f_{R}
	= 8\pi G\, T_{\mu\nu},
\end{equation}
where $f_{R} \equiv df/dR$ and $\Box = \nabla^\alpha \nabla_\alpha$.

We consider a spatially flat Friedmann–Robertson–Walker (FRW) metric in conformal time $\eta$,
\begin{equation}
	ds^{2} = a^{2}(\eta)\left[d\eta^{2} - \delta_{ij}dx^{i}dx^{j}\right],
\end{equation}
where $a(\eta)$ is the scale factor and $\mathcal{H} \equiv a'/a$ is the conformal Hubble parameter.

We assume that the matter content of the Universe can be modeled as a homogeneous and isotropic perfect cosmological fluid. The corresponding energy--momentum tensor is given by
\begin{equation}
	T_{\mu\nu}=(\rho+p)u_\mu u_\nu+p,g_{\mu\nu},
\end{equation}
where $u_\mu$ is the four-velocity of the comoving observer, while $\rho$ and $p$ denote the total energy density and pressure of the cosmic fluid, respectively.
\color{black}
The total energy density and pressure are defined as
\begin{equation}
	\rho \equiv \rho_b + \rho_c + \rho_r + \rho_\nu,
	\qquad
	p \equiv p_r + p_\nu,
\end{equation}
where baryons ($b$), cold dark matter ($c$), radiation ($r$), and massive neutrinos ($\nu$) are included. Cold dark matter and baryons are pressureless, while radiation and neutrinos contribute to the pressure sector.

Using the above metric, the modified Friedmann equations in metric $f(R)$ gravity can be written as
\begin{equation}
	3\mathcal{H}^{2} f_{R}
	= 8\pi G a^{2} \rho
	+ \frac{a^{2}}{2}\left(R f_{R} - f\right)
	- 3\mathcal{H} f_{R}',
\end{equation}

\begin{equation}
	\left(2\mathcal{H}' + \mathcal{H}^{2}\right) f_{R}
	= -8\pi G a^{2} p
	+ \frac{a^{2}}{2}\left(f - R f_{R}\right)
	+ \mathcal{H} f_{R}' + f_{R}''.
\end{equation}

Here a prime denotes differentiation with respect to conformal time $\eta$.

Assuming a barotropic equation of state,
\begin{equation}
	\bar{p}=w\bar{\rho},
\end{equation}
the conservation equation becomes
\begin{equation}
	\bar{\rho}' + 3(1+w)\mathcal{H}\bar{\rho}=0.
\end{equation}

Combining these relations leads to
\begin{equation}
	2(1+f_{R})(-\mathcal{H}'+\mathcal{H}^{2})
	+2\mathcal{H}f_{R}'-f_{R}''
	=
	8\pi G\,a^{2}\bar{\rho}(1+w).
\end{equation}

\subsection{Scalar Perturbations}

In this subsection, we briefly review the standard linear scalar perturbation formalism in metric $f(R)$ gravity following the well-established treatments presented in Refs.~\cite{Hwang1997,HwangNoh2001,Song2007,Tsujikawa2007,DeFelice2010,Sotiriou2010}. 
The purpose is to summarize the perturbation equations relevant to the subsequent cosmological analysis and to illustrate how modifications to the gravitational action influence the evolution of scalar modes associated with matter density and metric perturbations. Starting from the perturbed Friedmann--Robertson--Walker (FRW) metric, the first-order perturbation equations governing the gravitational potentials $\phi$ and $\psi$, the matter density contrast $\delta$, and the velocity potential $v$ are presented in their standard form. These equations constitute the theoretical basis for studying the growth of cosmic structures and for confronting modified gravity models with cosmological observations, including cosmic microwave background (CMB) anisotropies and large-scale structure measurements.
We specify that the Newtonian gauge is adopted for the study of linear scalar
perturbations.
We now consider linear scalar perturbations around the flat FRW background:
\begin{equation}
	ds^{2} = a^{2}(\eta)\left[(1 + 2\phi)d\eta^{2} - (1 - 2\psi)\,dx^{2}\right],
\end{equation}
where $\phi(\eta,x)$ and $\psi(\eta,x)$ are the scalar potentials.

The perturbed components of the energy--momentum tensor are given by
\begin{equation}
	\begin{split}
		\delta T^{0}_{0} &= \delta \rho = \rho_{0}\delta,\\
		\delta T^{0}_{i} &= -(1+\omega)\rho_{0}\partial_{i}v,\\
		\delta T^{i}_{j} &= -\delta p\,\delta^{i}_{j} + \Sigma^{i}_{j}
		= -\omega\rho_{0}\delta\,\delta^{i}_{j} + \Sigma^{i}_{j},
	\end{split}
\end{equation}
where $\Sigma^{i}_{j}$ denotes the anisotropic stress tensor satisfying
$\Sigma^{i}_{i}=0$.
where $v$ represents the velocity potential.

The first-order perturbed field equations, assuming the background equations hold, are given by
\begin{equation}
	(1 + f_{R})\,\delta G^{\mu}_{\nu}
	+ (R^{\mu}_{0\nu} + \nabla^{\mu}\nabla_{\nu} - \delta^{\mu}_{\nu}\Box) f_{RR}\,\delta R
	= 8\pi G\,\delta T^{\mu}_{\nu},
\end{equation}
where $f_{RR} \equiv \frac{d^{2}f}{dR^{2}}$.

Under these assumptions, the scalar perturbation equations reduce to
\begin{equation}
	\label{pert1} \phi-\psi = -\frac{f_{RR}}{1+f_{R}}\delta R,
\end{equation} 
which follows from the traceless spatial $(i\neq j)$ component of the linearized field equations. This equation explicitly demonstrates the emergence of gravitational slip in $f(R)$ gravity. Unlike standard Einstein gravity, where the absence of anisotropic stress implies $\phi=\psi$, the scalar mode associated with $f_{RR}$ acts as an effective anisotropic stress source generated purely by the modified gravitational sector. The perturbation of the Ricci scalar is given by 
\begin{equation} 
	\delta R = -\frac{2}{a^{2}} \left[ 3\psi'' + 6(\mathcal{H}'+\mathcal{H}^{2})\phi + 3\mathcal{H}(\phi'+3\psi') - k^{2}(\phi-2\psi) \right],
\end{equation} 
where primes denote derivatives with respect to conformal time and is the conformal Hubble parameter. The quantity $\delta R$ encodes the dynamical scalar degree of freedom intrinsic to metric $f(R)$ gravity and mediates the coupling between the modified gravity sector and the metric perturbations. The perturbed $(0,0)$ component of the modified Einstein equations yields 
\begin{equation}
	\begin{split} 
		\Big[ 3\mathcal{H}(\phi'+\psi') + k^{2}(\phi+\psi) + 3\mathcal{H}'\psi - (3\mathcal{H}'-6\mathcal{H}^{2})\phi \Big](1+f_{R}) \\ + (9\mathcal{H}\phi - 3\mathcal{H}\psi + 3\psi')f_{R}' = -a^{2}\kappa^{2}\rho_{0}\delta,
	\end{split}
\end{equation} 
which acts as the generalized Poisson equation in $f(R)$ gravity. In contrast to the standard Poisson equation of General Relativity, the gravitational potential is sourced not only by matter density perturbations but also by the dynamical evolution of the modified gravity scalar field through the terms proportional to $f_{R}$ and $f_{R}'$. Similarly, the trace of the spatial $(i,i)$ component leads to 
\begin{equation} 
	\begin{split} 
		\left[ \phi'' + \psi'' + 3\mathcal{H}(\phi'+\psi') + 3\mathcal{H}'\phi + (\mathcal{H}'+2\mathcal{H}^{2})\phi \right](1+f_{R}) \\ + (3\mathcal{H}\phi - \mathcal{H}\psi + 3\phi')f_{R}' + (3\phi-\psi)f_{R}'' = 0,
	\end{split} 
\end{equation} which governs the dynamical evolution of the scalar metric perturbations. The appearance of higher-order derivatives of $f_{R}$ reflects the existence of the extra propagating scalar mode characteristic of metric $f(R)$ theories. The perturbed momentum equation, corresponding to the $(0,i)$ component of the field equations, becomes 
\begin{equation}
	(2\phi-\psi)f_{R}' + (\phi'+\psi'+\mathcal{H}(\phi+\psi))(1+f_{R}) = -a^{2}\kappa^{2}\rho_{0}v,
\end{equation} 
where $v$ is the scalar velocity potential. This equation governs the evolution of matter velocity perturbations in the presence of modified gravity corrections. Finally, conservation of the total energy--momentum tensor,	$\nabla_{\mu}T^{\mu\nu}=0,$ leads to the standard continuity and Euler equations for matter perturbations, 
\begin{equation}
	\delta' - k^{2}v - 3\psi' = 0, \qquad \phi+\mathcal{H}v+v' = 0. 
\end{equation} 
These equations describe the evolution of matter density fluctuations and peculiar velocities within the perturbed spacetime geometry. Although the matter sector remains minimally coupled, its evolution is indirectly modified through the altered gravitational potentials generated by the additional scalar degree of freedom in the $f(R)$ sector. Collectively, the above equations form a closed system governing the linear evolution of scalar perturbations in interacting $f(R)$ cosmology. Their numerical implementation within the \texttt{hi\_class} framework enables a fully self-consistent computation of the CMB temperature and polarization anisotropies, matter power spectrum, weak-lensing observables, and structure-growth evolution in the presence of modified gravity and neutrino interactions.

\section{Perturbed $f(R)$ Gravity Coupled with Neutrinos}

In this section, we present the theoretical framework describing the coupling between neutrinos and the scalar degree of freedom in metric $f(R)$ gravity. The interaction is introduced through a phenomenological coupling term that modifies the conservation equation of the neutrino sector while preserving the covariant conservation of the total energy--momentum tensor. The resulting framework provides the basis for the background and perturbative cosmological analyses presented in the subsequent sections.

The action of the coupled system is
\begin{equation}
	S=\int d^{4}x\,\sqrt{-g}
	\left[
	\frac{f(R)}{16\pi G}
	+\mathcal{L}_{m}
	+\mathcal{L}_{\nu}
	+\mathcal{L}_{\mathrm{int}}
	\right],
\end{equation}
where $f(R)$ is a general function of the Ricci scalar $R$, $\mathcal{L}_{m}$ and $\mathcal{L}_{\nu}$ denote the matter and neutrino Lagrangians, respectively, while $\mathcal{L}_{\mathrm{int}}$ describes the interaction between the neutrino sector and the scalar degree of freedom associated with $f(R)$ gravity.

Varying the action with respect to the metric $g_{\mu\nu}$ yields the modified gravitational field equations
\begin{equation}
	f_RR_{\mu\nu}
	-\frac12f(R)g_{\mu\nu}
	+\left(g_{\mu\nu}\Box-\nabla_\mu\nabla_\nu\right)f_R
	=
	\kappa^{2}
	\left(
	T^{(m)}_{\mu\nu}
	+
	T^{(\nu)}_{\mu\nu}
	+
	T^{(\mathrm{int})}_{\mu\nu}
	\right),
\end{equation}
where $f_R\equiv df/dR$ and $\kappa^2 = 8\pi G$.

Taking the covariant divergence of both sides and using the contracted Bianchi identity gives
\begin{equation}
	\nabla^\mu
	\left(
	T^{(m)}_{\mu\nu}
	+
	T^{(\nu)}_{\mu\nu}
	+
	T^{(\mathrm{int})}_{\mu\nu}
	\right)
	=0.
\end{equation}

Assuming that the standard matter sector is separately conserved,
\begin{equation}
	\nabla^\mu T^{(m)}_{\mu\nu}=0,
\end{equation}
the conservation equation for the neutrino sector becomes
\begin{equation}
	\nabla^\mu T^{(\nu)}_{\mu\nu}
	=
	-
	\nabla^\mu
	T^{(\mathrm{int})}_{\mu\nu}.
\end{equation}

To close the system, we adopt the phenomenological ansatz
\begin{equation}
	T^{(\mathrm{int})}_{\mu\nu}
	=
	(1-f_R)
	T^{(\nu)}_{\mu\nu},
\end{equation}
which parametrizes an effective curvature-induced modification of the neutrino energy--momentum tensor through the scalar degree of freedom of $f(R)$ gravity.

Taking the covariant divergence of the interaction tensor yields
\begin{equation}
	\nabla^\mu
	T^{(\mathrm{int})}_{\mu\nu}
	=
	-
	(\nabla^\mu f_R)
	T^{(\nu)}_{\mu\nu}
	+
	(1-f_R)
	\nabla^\mu
	T^{(\nu)}_{\mu\nu}.
\end{equation}

Substituting this expression into the neutrino conservation equation gives
\begin{equation}
	\nabla^\mu
	T^{(\nu)}_{\mu\nu}
	=
	(\nabla^\mu f_R)
	T^{(\nu)}_{\mu\nu}
	-
	(1-f_R)
	\nabla^\mu
	T^{(\nu)}_{\mu\nu},
\end{equation}
which can be rearranged as
\begin{equation}
	(2-f_R)
	\nabla^\mu
	T^{(\nu)}_{\mu\nu}
	=
	(\nabla^\mu f_R)
	T^{(\nu)}_{\mu\nu}.
\end{equation}

Therefore,
\begin{equation}
	\nabla^\mu
	T^{(\nu)}_{\mu\nu}
	=
	\frac{
		(\nabla^\mu f_R)
		T^{(\nu)}_{\mu\nu}}
	{2-f_R}.
\end{equation}

Projecting this equation along the neutrino four-velocity $u^\nu$ leads to
\begin{equation}
	u^\nu
	\nabla^\mu
	T^{(\nu)}_{\mu\nu}
	=
	\frac{
		(u^\mu\nabla_\mu f_R)\rho_\nu
	}
	{2-f_R}
	\equiv
	-\Gamma\rho_\nu,
\end{equation}
where the effective interaction rate is defined by
\begin{equation}
	\Gamma
	\equiv
	-
	\frac{
		u^\mu\nabla_\mu f_R
	}
	{2-f_R}.
\end{equation}

The above definition provides a covariant parametrization of the energy exchange between neutrinos and the scalar degree of freedom in the adopted phenomenological framework. In the General Relativity limit, $f_R\rightarrow1$, the interaction rate vanishes, recovering the standard conservation law for neutrinos.

Expressing the evolution in terms of the e-folding variable $N=\ln a$, the modified neutrino continuity equation becomes
\begin{equation}
	\frac{d\rho_\nu}{dN}
	+
	3(1+w_\nu)\rho_\nu
	=
	-\Gamma\rho_\nu,
\end{equation}
where $w_\nu \equiv \frac{p_\nu}{\rho_\nu}$ is the neutrino equation-of-state parameter, with
$w_\nu \simeq 1/3$ in the relativistic regime and
$w_\nu \simeq 0$ after neutrinos become non-relativistic.

The above equations provide a self-consistent covariant description of the interaction between the neutrino sector and the scalar degree of freedom in metric $f(R)$ gravity. This formulation constitutes the theoretical basis for the cosmological background evolution and perturbation analyses presented in the following sections.

\subsection{Phenomenological treatment of the neutrino sector}

In the present analysis, the neutrino component is treated within the Boltzmann-moment (fluid) approximation, assuming a Fermi--Dirac background distribution and retaining the first few multipole moments of perturbations \cite{MaBertschinger1995,Lesgourgues2006,Blas2011}. This approach is used routinely in cosmological Boltzmann solvers such as \texttt{CLASS} and \texttt{hi\_class}, and it gives an accurate macroscopic description of relativistic and non-relativistic neutrino populations throughout cosmic history. Within this framework, we introduce an effective energy exchange between the dark sector and the neutrino fluid through a coupling term of the form
\begin{equation}
	Q_\nu = -\Gamma \rho_\nu,
\end{equation}
where $\Gamma$ denotes the phenomenological interaction rate. This term is the leading-order contribution in a derivative expansion of possible scalar--neutrino interactions, and it accounts for the dominant macroscopic effect of the coupling on the background and perturbation evolution.

The effective interaction is implemented directly at the background and linear perturbation levels within the \texttt{hi\_class} framework. The neutrino density contrast, velocity perturbation, and higher-order multipoles are evolved consistently with this modification throughout cosmic history.

$\Gamma$ is treated as a free parameter in the Bayesian fit rather than fixed directly by Eq.~(36). Across the dataset combinations considered in this work, the posterior constraints place $\Gamma$ in the range $0.05$--$0.07$ (Table~\ref{tab:combined_results}). This is comparable in order of magnitude to $df_R/dN$ evaluated directly from Eq.~(36) using the best-fit Hu--Sawicki parameters $C_1$, $C_2$, and $f_0$, shown in Fig.~\ref{fig:fr_coupling_triangle}. The two quantities are not expected to coincide exactly, since the fitted $\Gamma$ can absorb any curvature--neutrino coupling beyond the minimal form of Eq.~(26); their agreement in order of magnitude indicates that the interaction strength preferred by the data is consistent with what the fitted $f(R)$ dynamics alone would produce, rather than requiring an anomalously large or small coupling.

A derivation of $\Gamma$ from a specific microscopic Lagrangian is left for future work; this would require integrating out fermionic bilinears coupled to the scalar degree of freedom of $f(R)$ gravity. The phenomenological treatment adopted here satisfies covariant energy--momentum conservation and is sufficient to capture the main effect of a curvature--neutrino interaction on the observables considered in this work.

\subsection{Hu-Sawicki \( f(R) \) Gravity Model}
The functional form of the Hu--Sawicki model is given by Eq.(1). The first derivative of the Hu--Sawicki function with respect to the Ricci scalar is
\begin{equation}
	f_R
	\equiv
	\frac{df}{dR}
	=
	1
	-
	\frac{
		n c_1
		\left(
		\frac{R}{m^2}
		\right)^{n-1}
	}{
		\left[
		c_2
		\left(
		\frac{R}{m^2}
		\right)^n
		+
		1
		\right]^2
	}.
\end{equation}

Accordingly, the interaction parameter \(\Gamma\) becomes
\begin{equation}
	\Gamma
	\approx
	\frac{d f_R}{dN},
\end{equation}
where \(N=\ln a\) is the number of e-foldings.

In the high-curvature limit \(R \gg m^2\), the correction term becomes negligible and the model asymptotically approaches General Relativity,
\[
f(R)\rightarrow R.
\]
At low curvatures, however, the modification effectively mimics a cosmological constant and drives the late-time accelerated expansion of the Universe.

 The choice $n=4$ is motivated by the fact that larger values of $n$ lead to a steeper transition between the high-curvature regime (where General Relativity is recovered) and the low-curvature regime (where deviations from GR become dynamically relevant). Such behavior has been shown to improve the viability of $f(R)$ models in satisfying both cosmological and local gravity constraints, while still producing a sufficiently strong late-time acceleration effect. Similar choices of higher $n$ values have been explored in the literature as phenomenological extensions of the original Hu--Sawicki model~\cite{HuSawicki2007, Starobinsky2007, AmendolaTsujikawa2010,Luisa, YYY}.

\section{Numerical Analysis and Data Integration}
\label{sec:numerical_analysis}

To constrain the cosmological parameters of the interacting Hu--Sawicki $f(R)$ gravity model, we perform a Bayesian Markov Chain Monte Carlo (MCMC) analysis using the \texttt{MontePython} package~\cite{Audren2013MontePython, Brinckmann2019MontePython3}, interfaced with the modified-gravity Boltzmann solver \texttt{hi\_class}~\cite{Zumalacarregui2017}. The \texttt{hi\_class} framework extends the Einstein--Boltzmann solver \texttt{CLASS} to generalized scalar--tensor theories, evolving the full set of covariant background and perturbation equations directly from the underlying gravitational action rather than relying on phenomenological parameterizations.In this work, the Hu--Sawicki modified gravity sector and the effective neutrino--gravity interaction are implemented directly within \texttt{hi\_class}. At the background level, the interaction source term $Q_{\nu}$ modifies the standard neutrino continuity equation as  (Eq. (33)). This ensures the neutrino energy density evolves dynamically under the combined effects of cosmic expansion and curvature-induced energy exchange. Furthermore, $Q_{\nu}$ is similarly incorporated into the perturbed neutrino continuity and Euler equations. This comprehensive implementation allows for a self-consistent computation of the CMB temperature and polarization anisotropies, matter power spectrum, weak-lensing observables, and growth-rate evolution.

	At the perturbation level, the corresponding neutrino Euler and continuity equations are modified consistently by including the interaction source contributions in the momentum and density perturbation evolution equations. In synchronous gauge, the exact perturbed continuity equation is given by
\begin{equation}
	\dot{\delta}_{\nu}
	+ (1+w_{\nu})
	\left(
	\theta_{\nu}+\frac{\dot{h}}{2}
	\right)
	+3H\left(\omega-w_{\nu}\right)\delta_{\nu}
	=
	-\Gamma \delta_{\nu},
\end{equation}
while the modified Euler equation becomes
\begin{equation}
	\dot{\theta}_{\nu}
	+
	H(1-3w_{\nu})\theta_{\nu}
	-
	\frac{k^2 \omega}{1+w_{\nu}}\delta_{\nu}
	=
	-\Gamma \theta_{\nu}.
\end{equation}

Here, $\delta_{\nu}$ and $\theta_{\nu}$ denote the neutrino density contrast and velocity divergence, respectively. These equations represent the exact macroscopic energy and momentum conservation laws for the interacting neutrino fluid.

The implementation is performed directly within the neutrino perturbation modules of \texttt{hi\_class}. Rather than relying on approximate illustrative models, the interaction source terms are rigorously added to both the background evolution routines and the exact linearized Boltzmann multipole hierarchy. Specifically, these macroscopic interaction terms explicitly modify the monopole ($\ell=0$) and dipole ($\ell=1$) evolution equations. The modified equations are evolved simultaneously with the Horndeski scalar--tensor sector, ensuring full covariance, numerical stability, and a completely self-consistent computation of the perturbation dynamics across all relevant physical scales throughout the cosmological evolution.

To maintain numerical stability and avoid spurious higher-order truncation effects, the neutrino hierarchy is evolved using the standard Boltzmann closure approximation already implemented in \texttt{hi\_class}/\texttt{CLASS}. No additional phenomenological closure relation is introduced beyond the standard truncation procedure employed for relativistic species in linear perturbation theory.

The numerical evolution of perturbations is restricted to the linear regime, where the modified gravity corrections remain perturbatively well-defined and numerically stable. Throughout the analysis, the parameter space is constrained to satisfy the standard viability conditions of \(f(R)\) gravity,
\begin{equation}
	f_R > 0,
	\qquad
	f_{RR} > 0,
\end{equation}
which ensure a positive effective gravitational coupling and a positive scalaron mass squared, thereby avoiding ghost instabilities, tachyonic modes, and pathological scalar dynamics.

To obtain robust cosmological constraints and efficiently break parameter degeneracies, we combine several complementary cosmological probes spanning both early- and late-time epochs of cosmic evolution.

\subsection{Likelihood Construction in \texttt{MontePython}}
\label{sec:montepython_likelihoods}

The total likelihood function is constructed as the product of the individual likelihoods associated with each cosmological probe,
\begin{equation}
	\mathcal{L}(\mathrm{data}|\theta)
	=
	\mathcal{L}_{\mathrm{CMB}}
	\,
	\mathcal{L}_{\mathrm{lensing}}
	\,
	\mathcal{L}_{\mathrm{BAO}}
	\,
	\mathcal{L}_{\mathrm{CC}}
	\,
	\mathcal{L}_{\mathrm{SN}},
\end{equation}
where \(\theta\) denotes the full vector of cosmological and modified-gravity parameters. The posterior probability distribution is then given by
\begin{equation}
	\mathcal{P}(\theta|\mathrm{data})
	\propto
	\mathcal{L}(\mathrm{data}|\theta)\,
	\pi(\theta),
\end{equation}
with \(\pi(\theta)\) representing the prior distributions.

\subsubsection{CMB and CMB Lensing Likelihoods}

We employ the official \textit{Planck} 2018 temperature and polarization likelihoods (plikTTTEEE+lowl+lowE)~\cite{Planck2018}, which provide stringent constraints on both the cosmological background and the evolution of linear perturbations. In addition, we include the CMB lensing reconstruction likelihood~\cite{Aghanim1}, derived from the four-point correlation function of the CMB anisotropies, which probes the integrated matter distribution along the line of sight and is particularly sensitive to modifications of gravity and neutrino physics.

The CMB likelihood is evaluated as
\begin{equation}
	\mathcal{L}_{\mathrm{CMB}}
	\propto
	\exp\left[
	-\frac{1}{2}
	\Delta \mathbf{C}_{\ell}^{\mathrm{T}}
	\mathbf{C}_{\mathrm{CMB}}^{-1}
	\Delta \mathbf{C}_{\ell}
	\right],
\end{equation}
where
\begin{equation}
	\Delta \mathbf{C}_{\ell}
	=
	\mathbf{C}_{\ell}^{\mathrm{obs}}
	-
	\mathbf{C}_{\ell}^{\mathrm{theory}},
\end{equation}
and \(\mathbf{C}_{\mathrm{CMB}}\) denotes the covariance matrix of the observed angular power spectra.

\subsubsection{Type Ia Supernovae (Pantheon+)}

To constrain the late-time expansion history, we employ the Pantheon+ compilation~\cite{42, ss}, consisting of 1701 spectroscopically confirmed Type Ia supernovae spanning the redshift range
\[
0.001 < z < 2.3.
\]
The Pantheon+ sample provides one of the most precise measurements of the luminosity distance relation and therefore strongly constrains the background cosmological dynamics.

The supernova likelihood function is written as
\begin{equation}
	\mathcal{L}_{\mathrm{SN}}
	\propto
	\exp\left[
	-\frac{1}{2}
	\sum_j
	\frac{
		\left(
		\mu_j^{\mathrm{obs}}
		-
		\mu_j^{\mathrm{model}}
		\right)^2
	}{
		\sigma_{\mu,j}^{2}
	}
	\right],
\end{equation}
where \(\mu_j^{\mathrm{obs}}\) and \(\mu_j^{\mathrm{model}}\) denote the observed and theoretical distance moduli, respectively, while \(\sigma_{\mu,j}\) represents the associated observational uncertainty.

\subsubsection{Baryon Acoustic Oscillations (BAO)}

Baryon Acoustic Oscillations (BAO) constitute one of the most robust geometrical probes of the late-time expansion history of the Universe. The characteristic BAO scale originates from sound waves propagating in the tightly coupled photon--baryon plasma prior to recombination and acts as a standard ruler calibrated by early-Universe physics.

In this work, we use the latest BAO measurements from the \textit{Dark Energy Spectroscopic Instrument} (DESI) Data Release 2 (DR2)~\cite{DESIDR2_1,DESIDR2_2}, representing the most precise three-dimensional mapping of large-scale structure currently available. The DESI DR2 dataset spans a broad redshift range, including low-redshift galaxy samples, luminous red galaxies, quasars, and Ly\(\alpha\)-forest measurements, thereby providing powerful constraints on both the transverse and radial expansion history.

The relevant BAO observables employed in the analysis are
\begin{equation}
	\frac{D_M(z)}{r_d},
	\qquad
	\frac{D_H(z)}{r_d}
	=
	\frac{c}{H(z)\,r_d},
\end{equation}
where \(D_M(z)\) denotes the comoving angular diameter distance, \(D_H(z)\) is the Hubble distance, and \(r_d\) represents the comoving sound horizon at the baryon drag epoch.

The comoving angular diameter distance is defined by
\begin{equation}
	D_M(z)
	=
	c
	\int_0^z
	\frac{dz'}{H(z')}.
\end{equation}

The BAO likelihood function is constructed using the full covariance matrix released by the DESI collaboration, accounting for correlations among the different observables and redshift bins,
\begin{equation}
	\mathcal{L}_{\mathrm{BAO}}
	\propto
	\exp\left[
	-\frac{1}{2}
	\Delta \mathbf{D}^{\mathrm{T}}
	\mathbf{C}_{\mathrm{BAO}}^{-1}
	\Delta \mathbf{D}
	\right],
\end{equation}
where \(\Delta \mathbf{D}\) denotes the vector of differences between the observed and model-predicted BAO distance ratios.

The DESI DR2 BAO dataset employed in this analysis covers effective redshifts in the approximate range
\[
z \sim 0.3 - 2.3,
\]
including both transverse \((D_M/r_d)\) and radial \((D_H/r_d)\) measurements where available.

\subsubsection{Cosmic Chronometers (CC)}

We further include 32 measurements of the Hubble parameter \(H(z)\) obtained from the cosmic chronometer technique~\cite{Ratra1, Aghanim, Di Valentino1, Cardona2023}, covering the redshift interval
\[
0.07 < z < 1.965.
\]
Unlike distance-based probes, cosmic chronometers directly estimate the expansion rate through the differential age evolution of passively evolving galaxies, making them largely independent of any assumed cosmological model.

The corresponding likelihood is expressed as
\begin{equation}
	\mathcal{L}_{\mathrm{CC}}
	\propto
	\exp\left[
	-\frac{1}{2}
	\Delta \mathbf{H}^{\mathrm{T}}
	\mathbf{C}_{\mathrm{CC}}^{-1}
	\Delta \mathbf{H}
	\right],
\end{equation}
where
\begin{equation}
	\Delta \mathbf{H}
	=
	\mathbf{H}_{\mathrm{obs}}
	-
	\mathbf{H}_{\mathrm{model}}.
\end{equation}

\subsection{Priors and Sampling Strategy}

Flat priors are adopted for both the standard cosmological parameters and the modified-gravity parameters following Refs.~\cite{Ratra1, Aghanim, Aghanim1, Di Valentino1, Cardona2023}. The adopted parameter ranges are summarized in Table~\ref{tab:priors}. The priors associated with the modified-gravity sector are chosen to ensure numerical stability, consistency with viable Hu--Sawicki cosmologies, and compatibility with local gravity constraints.

Sampling is performed using the Metropolis--Hastings algorithm with adaptive proposal covariance matrices. Convergence of the Markov chains is assessed using the Gelman--Rubin criterion,
\[
R-1 < 0.01,
\]
while maintaining acceptable sampling efficiency throughout the numerical analysis. The best-fit parameters correspond to the minimum of the total chi-squared function,
\begin{equation}
	\chi^2_{\mathrm{total}}
	=
	\sum_i
	-2\ln\mathcal{L}_i,
\end{equation}
whereas the marginalized posterior distributions provide the associated credible intervals.

\begin{table}[H]
	\centering
	\caption{Flat priors adopted for the cosmological and modified-gravity parameters used in the MCMC analysis.}
	\label{tab:priors}
	\resizebox{0.34\textwidth}{!}{
		\begin{tabular}{|c|c|}
			\hline
			Parameter & Prior Range \\
			\hline
			
			$\Omega_b h^2$ & [0.005, 0.10] \\
			\hline
			
			$\Omega_c h^2$ & [0.001, 0.20] \\
			\hline
			
			$\tau$ & [0.01, 0.80] \\
			\hline
			
			$n_s$ & [0.80, 1.20] \\
			\hline
			
			$\log(10^{10} A_s)$ & [1.60, 3.90] \\
			\hline
			
			$100\,\theta_{\rm MC}$ & [0.50, 10.50] \\
			\hline
			
			$C_1~(\times10^{-3})$ & [0.50, 2.50] \\
			\hline
			
			$C_2~(\times10^{-5})$ & [1.00, 10.00] \\
			\hline
			
			$f_0 \equiv f_R(z=0)$ & [0.90, 1.10] \\
			\hline
			
			$\Gamma$ & [0.00, 0.50] \\
			\hline
			
			$H_0~(\mathrm{km\,s^{-1}\,Mpc^{-1}})$ & [50, 90] \\
			\hline
			
			$\Omega_\nu$ & [0.00, 0.10] \\
			\hline
			
		\end{tabular}
	}
\end{table}

By jointly analyzing these complementary cosmological probes within the \texttt{MontePython}--\texttt{hi\_class} framework, we obtain a statistically robust and self-consistent inference of the cosmological and modified-gravity parameters. The combined multi-probe analysis enables a stringent observational test of the interacting Hu--Sawicki \(f(R)\) gravity model and its implications for late-time cosmic acceleration, large-scale structure formation, and the cosmological tension problem.


\section{Results and Discussions}

To investigate the impact of modified gravity and neutrino--gravity interactions on cosmological parameter inference, we divide the observational data into two complementary combinations. The first dataset, DC1, consists of late-time probes ($\mathrm{CC+SN}$), while the second dataset, DC2, includes early- and intermediate-time observations ($\mathrm{BAO+CMB+Lensing}$). This separation allows us to examine the extent to which the inferred cosmological parameters depend on the underlying observational regime and to assess the consistency between late- and early-Universe constraints. In addition, we perform a joint analysis using the full dataset in order to obtain the most stringent parameter bounds and to evaluate the overall viability of the models. Within this framework, we compare the flat $\Lambda$CDM model, the non-interacting Hu--Sawicki $f(R)$ gravity scenario, and the interacting Hu--Sawicki $f(R)$ model. Particular attention is devoted to the Hubble constant $H_0$, the matter density parameter $\Omega_m$, the clustering amplitude $S_8$, and the summed neutrino mass $\Sigma m_\nu$, as these quantities are especially sensitive to modifications of the gravitational sector and are closely connected to current cosmological tensions. Beyond the comparison of parameter constraints, we also examine the statistical performance of each model through the corresponding goodness-of-fit measures. This analysis provides a unified framework for assessing whether modified gravity and neutrino interactions can alleviate existing cosmological tensions while remaining consistent with the full range of available observational data.
\subsection{Cosmological constraints and model comparison}

The cosmological constraints obtained for the flat $\Lambda$CDM model, the noninteracting Hu--Sawicki $f(R)$ scenario, and the interacting Hu--Sawicki $f(R)$ framework are summarized in Table~\ref{tab:combined_results}. 

A common feature among all three models is the shift in the inferred Hubble constant when early-Universe observations are included. For the $\Lambda$CDM model, the preferred value decreases from $H_0=71.21\pm1.68~\mathrm{km\,s^{-1}\,Mpc^{-1}}$ for DC1 to $H_0=68.42\pm1.42~\mathrm{km\,s^{-1}\,Mpc^{-1}}$ for DC2. A similar trend is observed in the modified-gravity scenarios, with the noninteracting Hu--Sawicki model yielding $H_0=71.14\pm1.81~\mathrm{km\,s^{-1}\,Mpc^{-1}}$ (DC1) and $H_0=69.25\pm1.42~\mathrm{km\,s^{-1}\,Mpc^{-1}}$ (DC2), while the interacting model gives $H_0=70.92\pm1.92~\mathrm{km\,s^{-1}\,Mpc^{-1}}$ and $H_0=69.16\pm1.75~\mathrm{km\,s^{-1}\,Mpc^{-1}}$, respectively. This behavior reflects the well-known tendency of early-Universe probes to favor lower expansion rates compared with late-time observations.

The matter density parameter exhibits a corresponding evolution between the two dataset combinations. In the $\Lambda$CDM case, $\Omega_m$ increases from $0.297\pm0.009$ to $0.303\pm0.009$ when moving from DC1 to DC2. The modified-gravity scenarios show a similar behavior, although the interacting model retains a comparatively lower value, $\Omega_m=0.296\pm0.008$, for DC2. This result suggests that the additional degrees of freedom introduced by modified gravity and the interaction sector can partially alter the preferred matter content while remaining consistent with observational constraints.

In contrast to the noticeable shifts in $H_0$ and $\Omega_m$, the clustering parameter $S_8$ remains remarkably stable across all models and dataset combinations. The inferred values range from $S_8=0.797\pm0.018$ to $S_8=0.809\pm0.013$, indicating that all three cosmological scenarios provide a comparable description of the observed growth of large-scale structures. The interacting model yields the lowest central values of $S_8$, although the differences relative to the other scenarios remain within the quoted uncertainties.

The inclusion of BAO, CMB, and lensing information substantially improves the constraints on the parameters associated with the modified-gravity sector. For both Hu--Sawicki scenarios, the allowed regions of the parameters $C_1$, $C_2$, and $f_0$ become noticeably narrower under DC2, reflecting the greater constraining power of early-Universe observations on departures from General Relativity. In the interacting framework, meaningful limits are also obtained for the coupling parameter $\Gamma$ and the summed neutrino mass, with $\Gamma=0.0645\pm0.034$ and $\Sigma m_\nu<0.24~\mathrm{eV}$ for DC1, improving to $\Gamma=0.069\pm0.017$ and $\Sigma m_\nu<0.18~\mathrm{eV}$ for DC2.

\begin{figure}[H]
	\centering
	\includegraphics[width=\columnwidth]{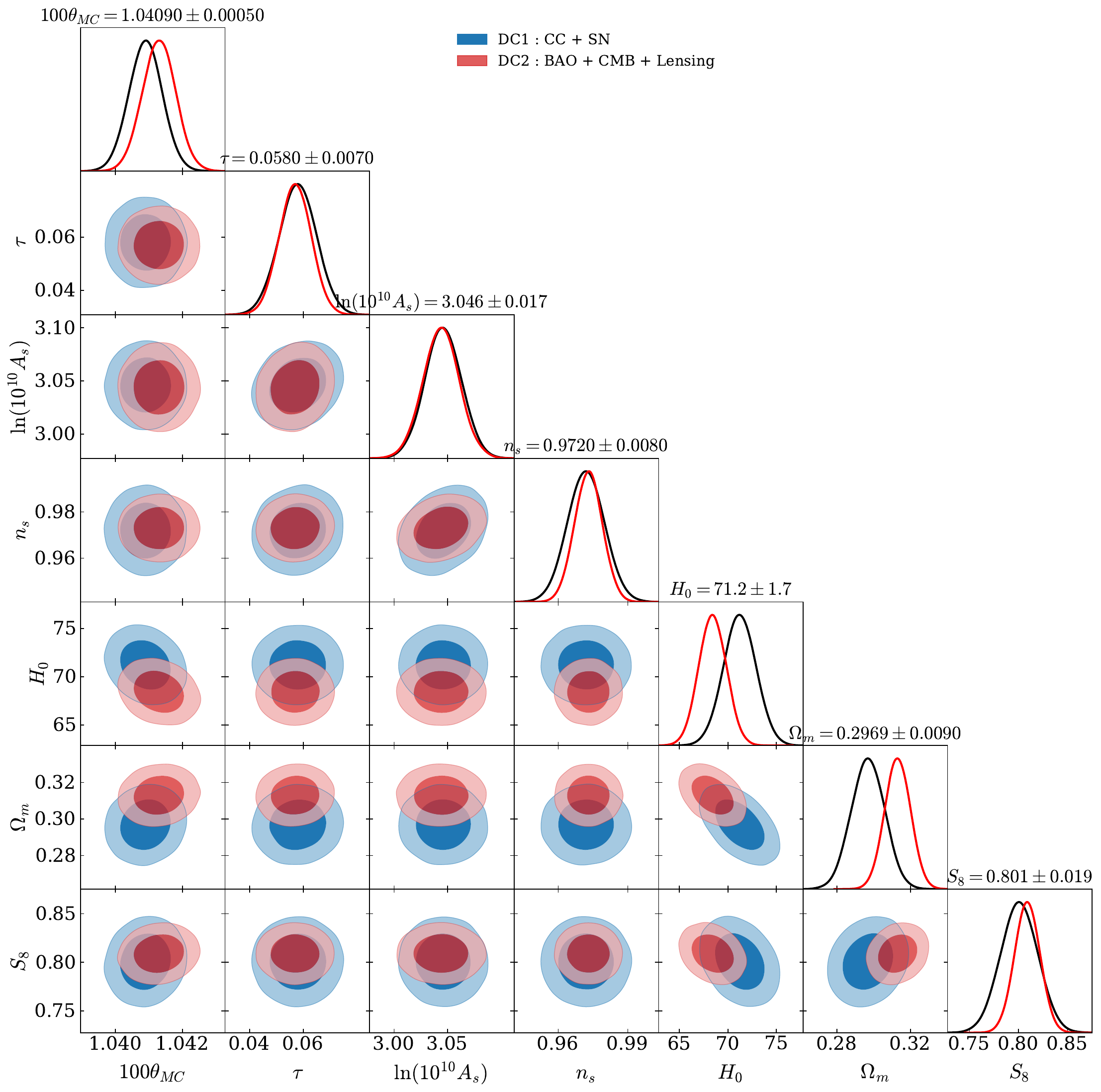}
	\caption{Joint posterior distributions and parameter correlations for the flat $\Lambda$CDM model using the DC1 ($\mathrm{CC+SN}$) and DC2 ($\mathrm{BAO+CMB+Lensing}$) dataset combinations.}
	\label{fig:lcdm_triangle}
\end{figure}

\begin{figure}[H]
	\centering
	\includegraphics[width=\columnwidth]{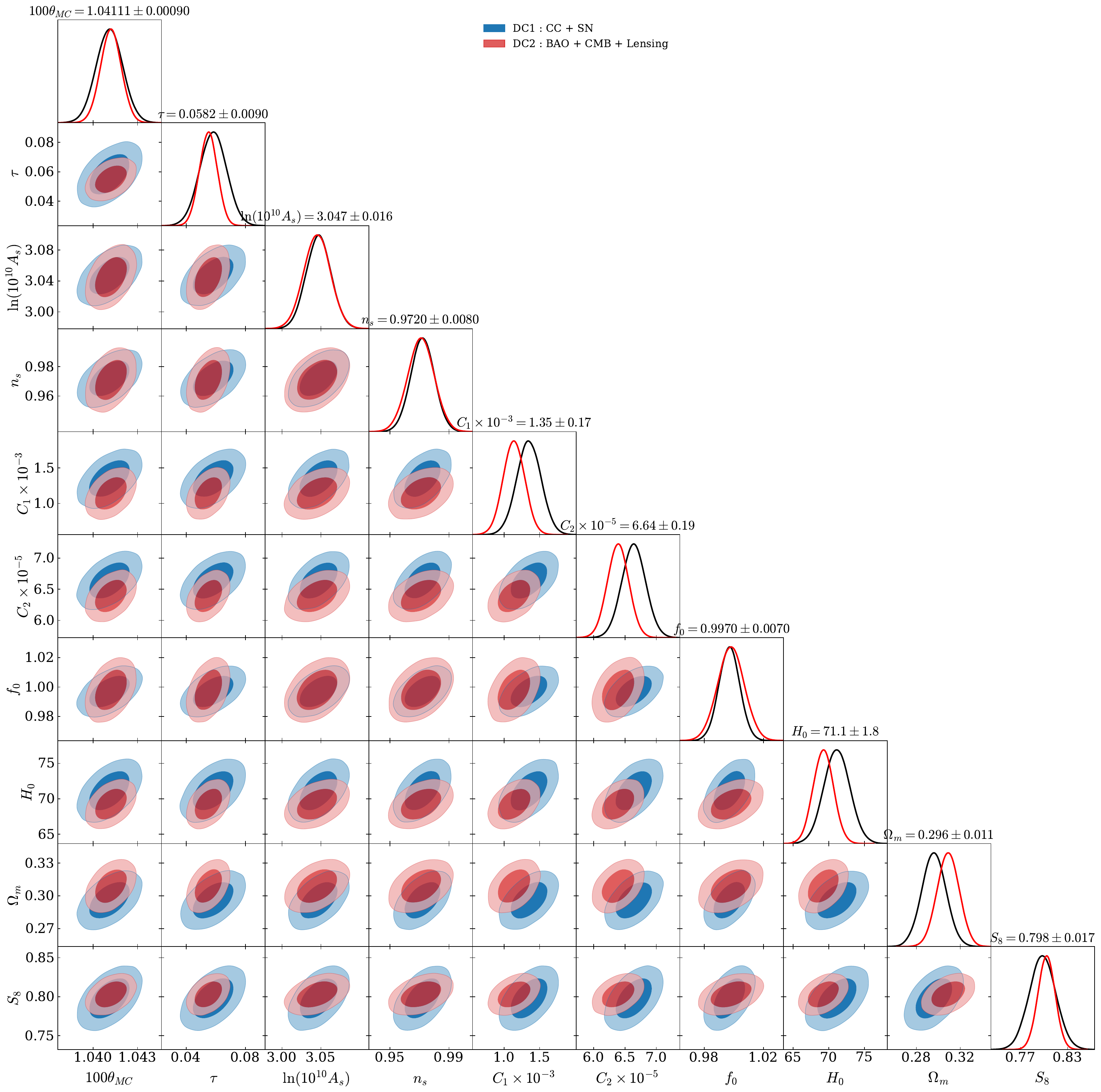}
	\caption{Joint posterior distributions and parameter correlations for the Hu--Sawicki $f(R)$ model without neutrino coupling using the DC1 ($\mathrm{CC+SN}$) and DC2 ($\mathrm{BAO+CMB+Lensing}$) dataset combinations.}
	\label{fig:fr_nocoupling_triangle}
\end{figure}

\begin{figure}[H]
	\centering
	\includegraphics[width=\columnwidth]{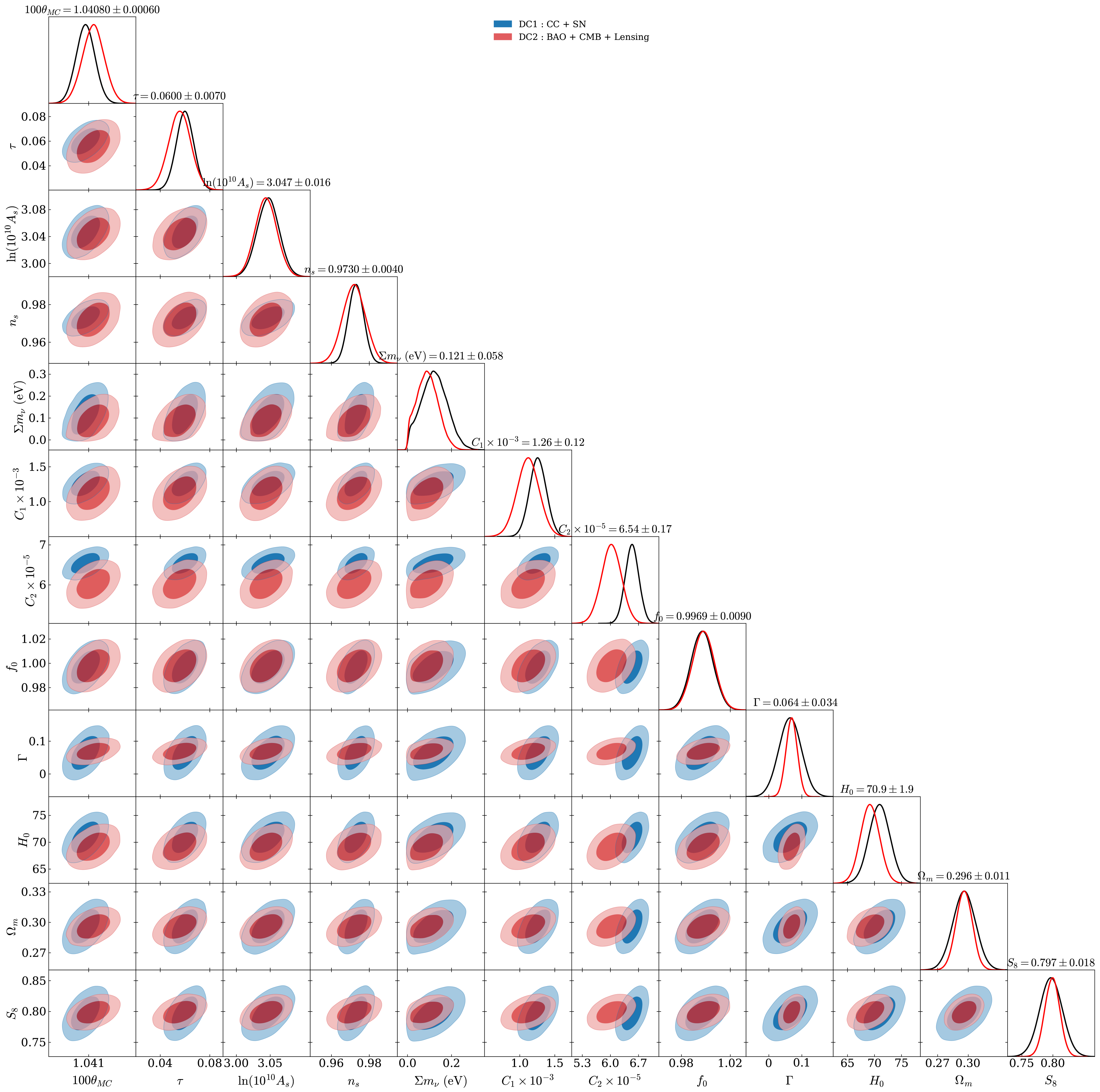}
	\caption{Joint posterior distributions and parameter correlations for the interacting Hu--Sawicki $f(R)$ model with neutrino coupling using the DC1 ($\mathrm{CC+SN}$) and DC2 ($\mathrm{BAO+CMB+Lensing}$) dataset combinations.}
	\label{fig:fr_coupling_triangle}
\end{figure}

The combined analysis shows that the Hu--Sawicki extensions remain fully compatible with current cosmological observations while providing additional flexibility in the description of the expansion history and structure-growth evolution. The posterior distributions displayed in Figs.~\ref{fig:lcdm_triangle}, \ref{fig:fr_nocoupling_triangle}, and \ref{fig:fr_coupling_triangle} further illustrate the parameter degeneracies and the impact of the different dataset combinations on the inferred cosmological constraints.

\begin{table}[H]
	\centering
	\caption{Cosmological parameter constraints for $\Lambda$CDM, Hu--Sawicki $f(R)$ (non-interacting), and Interacting Hu--Sawicki $f(R)$ models using DC1 and DC2 datasets.}
	\label{tab:combined_results}
	\resizebox{\textwidth}{!}{
		\begin{tabular}{l cc| cc| cc}
			\toprule
			& \multicolumn{2}{c}{\textbf{$\Lambda$CDM}} & \multicolumn{2}{c}{\textbf{$f(R)$ (Non-Int)}} & \multicolumn{2}{c}{\textbf{Interacting $f(R)$}} \\
			\cmidrule(lr){2-3} \cmidrule(lr){4-5} \cmidrule(lr){6-7}
			\textbf{Parameter} & \textbf{DC1} & \textbf{DC2} & \textbf{DC1} & \textbf{DC2} & \textbf{DC1} & \textbf{DC2} \\
			\midrule
			\hline
			$\Omega_b h^2$ & $0.02232 \pm 0.00023$ & $0.02234 \pm 0.00020$ & $0.02227 \pm 0.00021$ & $0.02234 \pm 0.00018$ & $0.02227 \pm 0.00018$ & $0.02232 \pm 0.00019$ \\
			$\Omega_c h^2$ & $0.1197 \pm 0.0037$ & $0.1196 \pm 0.0032$ & $0.1195 \pm 0.0042$ & $0.1192 \pm 0.0035$ & $0.1191 \pm 0.0030$ & $0.1192 \pm 0.0036$ \\
			$100\theta_{MC}$ & $1.0409 \pm 0.0005$ & $1.0413 \pm 0.0005$ & $1.0411 \pm 0.0009$ & $1.0412 \pm 0.0007$ & $1.0408 \pm 0.0006$ & $1.0413 \pm 0.0007$ \\
			$\tau$ & $0.058 \pm 0.007$ & $0.057 \pm 0.006$ & $0.0582 \pm 0.009$ & $0.055 \pm 0.006$ & $0.060 \pm 0.007$ & $0.0557 \pm 0.009$ \\
			$\ln(10^{10}A_s)$ & $3.046 \pm 0.017$ & $3.044 \pm 0.017$ & $3.047 \pm 0.016$ & $3.045 \pm 0.017$ & $3.047 \pm 0.016$ & $3.044 \pm 0.016$ \\
			$n_s$ & $0.972 \pm 0.008$ & $0.973 \pm 0.006$ & $0.972 \pm 0.008$ & $0.971 \pm 0.009$ & $0.973 \pm 0.004$ & $0.972 \pm 0.006$ \\
			$\Sigma m_\nu$ (eV) & - & - & - & - & $< 0.24$ & $< 0.18$ \\
			$C_1 (\times 10^{-3})$ & - & - & $1.35 \pm 0.17$ & $1.14 \pm 0.15$ & $1.26 \pm 0.12$ & $1.12 \pm 0.16$ \\
			$C_2 (\times 10^{-5})$ & - & - & $6.64 \pm 0.19$ & $6.39 \pm 0.17$ & $6.54 \pm 0.17$ & $6.026 \pm 0.25$ \\
			$f_0$ & - & - & $0.997 \pm 0.007$ & $0.998 \pm 0.009$ & $0.997 \pm 0.009$ & $0.998 \pm 0.009$ \\
			$\Gamma$ & - & - & - & - & $0.0645 \pm 0.034$ & $0.069 \pm 0.017$ \\
			$H_0$ & $71.21 \pm 1.68$ & $68.42 \pm 1.42$ & $71.14 \pm 1.81$ & $69.25 \pm 1.42$ & $70.92 \pm 1.92$ & $69.16 \pm 1.75$ \\
			$\Omega_m$ & $0.297 \pm 0.009$ & $0.313 \pm 0.007$ & $0.296 \pm 0.011$ & $0.309 \pm 0.010$ & $0.296 \pm 0.011$ & $0.296 \pm 0.008$ \\
			$S_8$ & $0.801 \pm 0.019$ & $0.809 \pm 0.013$ & $0.798 \pm 0.017$ & $0.803 \pm 0.011$ & $0.797 \pm 0.018$ & $0.799 \pm 0.012$ \\
			\midrule
			$\chi^2_{\rm min}$ & $1551.4$ & $2804.9$ & $1531.7$ & $2794.5$ & $1526.5$ & $2787.8$ \\
			\bottomrule
	\end{tabular}}
\end{table}

\subsection*{Cosmological Tensions from the Full Dataset Combination: $\Lambda$CDM, Hu--Sawicki $f(R)$, and Interacting $f(R)$ Gravity}
\label{sec:results}

To evaluate the observational viability of our proposed models and their capacity to alleviate cosmological tensions, we present a joint analysis of the standard $\Lambda$CDM, the non-interacting Hu--Sawicki $f(R)$, and the interacting Hu--Sawicki $f(R)$ frameworks. The marginalized parameter constraints for all models across various dataset combinations are summarized in Table~\ref{tab:combined_cosmo_models}.

\begin{figure}[H]
	\centering
	\includegraphics[width=\textwidth]{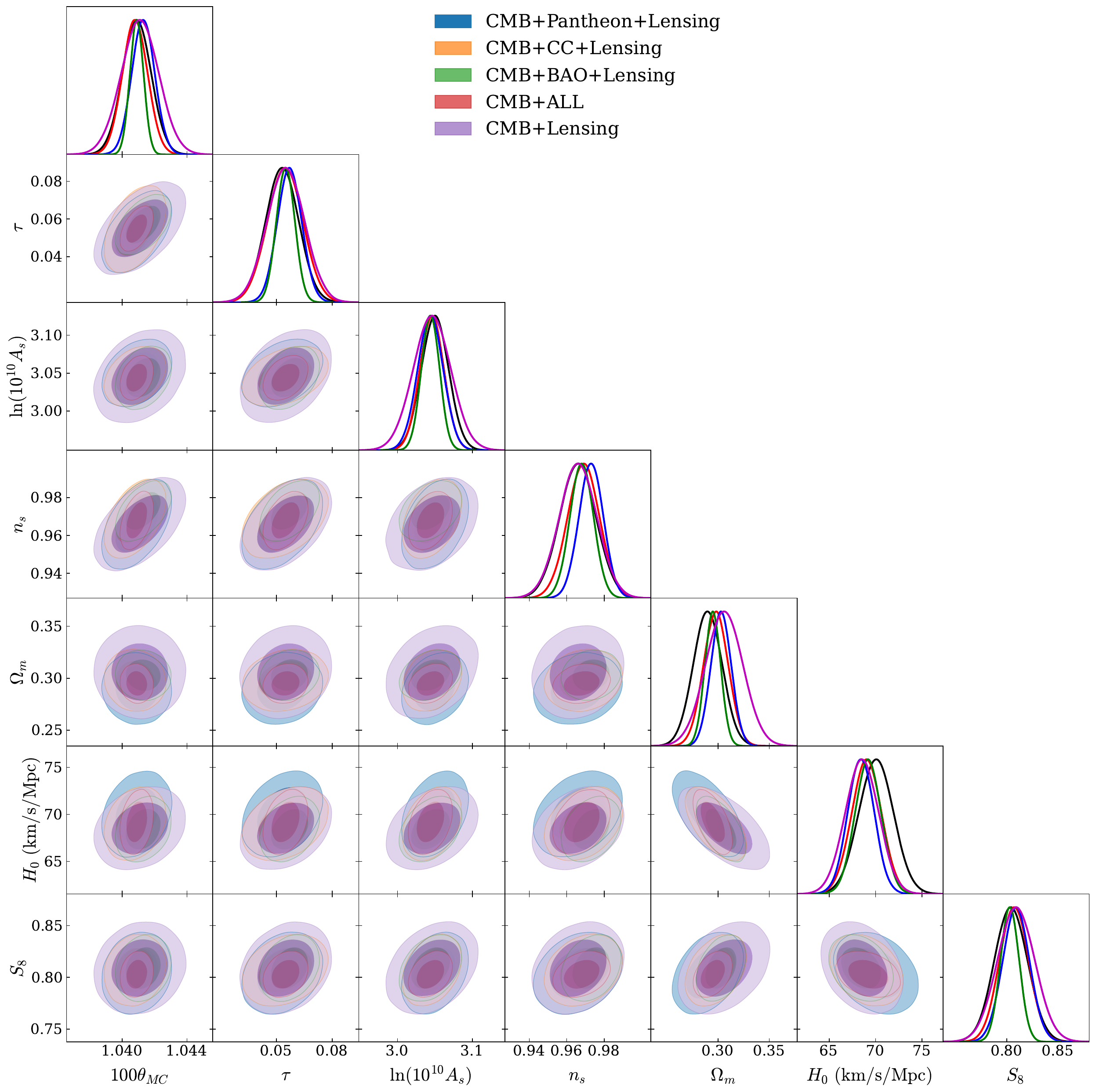}
	\caption{Triangle plot of the marginalized posterior distributions for the $\Lambda$CDM model parameters using different combinations of CMB, Pantheon, CC, BAO, and lensing datasets. The filled contours correspond to the $1\sigma$ and $2\sigma$ confidence regions.}
	\label{fig:LCDM_triangle}
\end{figure}

The baseline $\Lambda$CDM model yields an intermediate expansion rate of $H_0 = 69.19 \pm 1.42$~km~s$^{-1}$Mpc$^{-1}$ and a clustering amplitude of $S_8 = 0.803 \pm 0.009$ for the full \textbf{CMB+ALL} dataset (see Fig.~\ref{fig:LCDM_triangle} for the posterior distributions). Late-time distance probes pull $H_0$ in different directions: including the Pantheon sample raises it to $H_0 = 70.06^{+1.85}_{-1.87}$~km~s$^{-1}$Mpc$^{-1}$, while Cosmic Chronometers alone give a value close to the full-dataset result ($H_0 = 69.03^{+1.6}_{-1.6}$~km~s$^{-1}$Mpc$^{-1}$), and the acoustic-scale information from BAO favors the lowest value among all combinations. Despite the model's overall success in describing large-scale structure observations, these results confirm that the well-known $H_0$ and $S_8$ tensions remain unresolved within the standard framework.

\begin{figure}[H]
	\centering
	\includegraphics[width=\columnwidth]{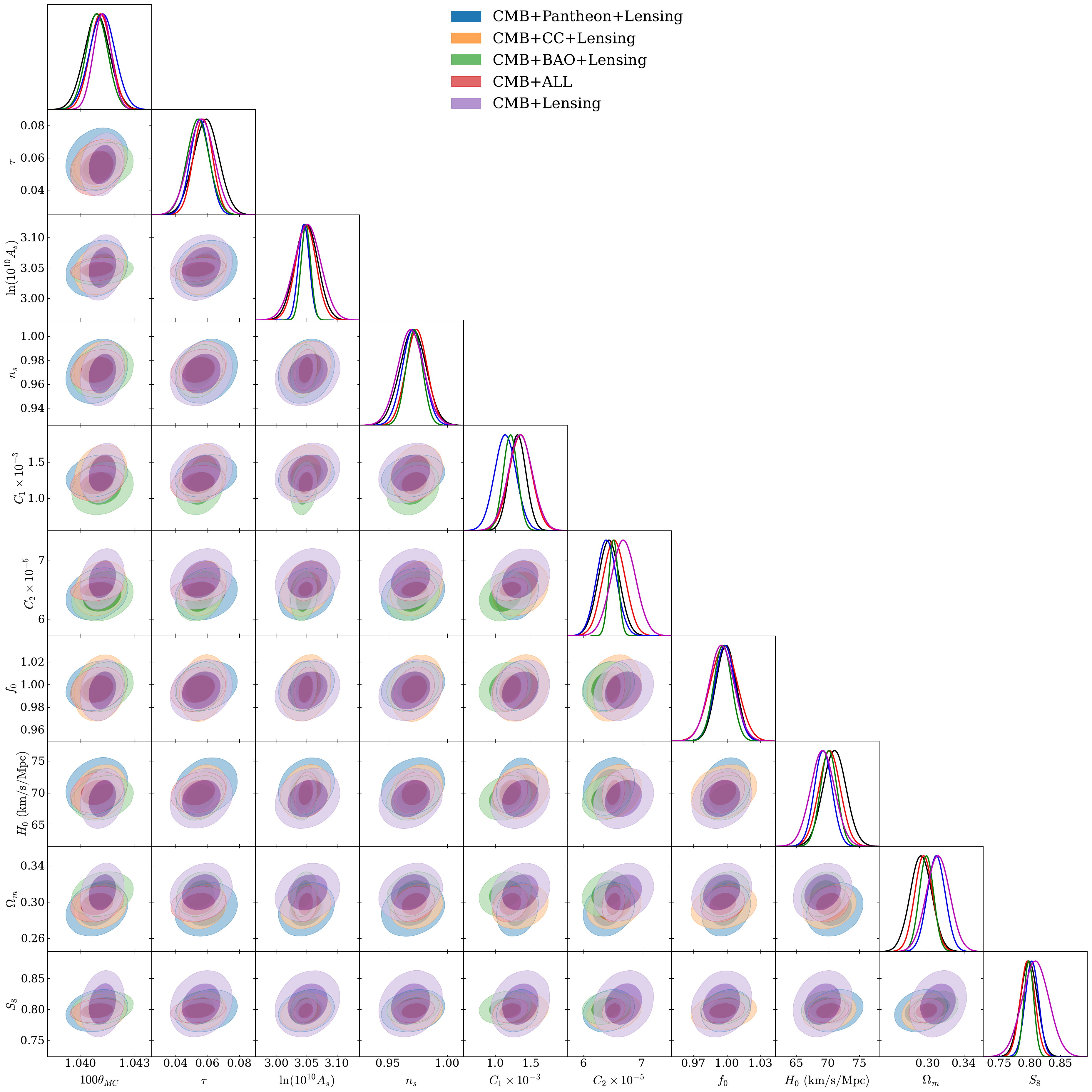}
	\caption{
		Triangle plot showing the marginalized posterior distributions and confidence contours for the cosmological and Hu--Sawicki $f(R)$ gravity parameters obtained from the combined \textbf{CMB+ALL} dataset analysis at the $68\%$ confidence levels. 
	}
	\label{fig:4}
\end{figure}

Moving beyond General Relativity, the non-interacting Hu--Sawicki $f(R)$ modification primarily alters late-time dynamics while preserving the early-Universe thermodynamic history; baseline parameters such as $\Omega_b h^2$ and $n_s$ remain virtually unchanged across all datasets. As illustrated in Fig.~\ref{fig:4}, the modified gravity parameters are constrained to $C_1 \sim \mathcal{O}(10^{-3})$, $C_2 \sim \mathcal{O}(10^{-5})$, and $f_0 \simeq 1$, indicating mild departures from GR. Phenomenologically, this geometric modification simultaneously elevates the expansion rate ($H_0 \sim 69$--$71$~km~s$^{-1}$Mpc$^{-1}$) and moderates the clustering amplitude ($S_8 \simeq 0.80$), thereby reducing discrepancies with local distance-ladder and weak-lensing measurements. The improvement in $\chi^2_{\rm min}$ is large enough that the non-interacting Hu--Sawicki model is preferred over $\Lambda$CDM by both the AIC and the BIC (Table~\ref{tab:AIC_BIC_models0}), although the stronger complexity penalty of the BIC leaves the two models much closer to each other than the AIC comparison alone would suggest.

\begin{figure}[H]
	\centering
	\includegraphics[width=\columnwidth]{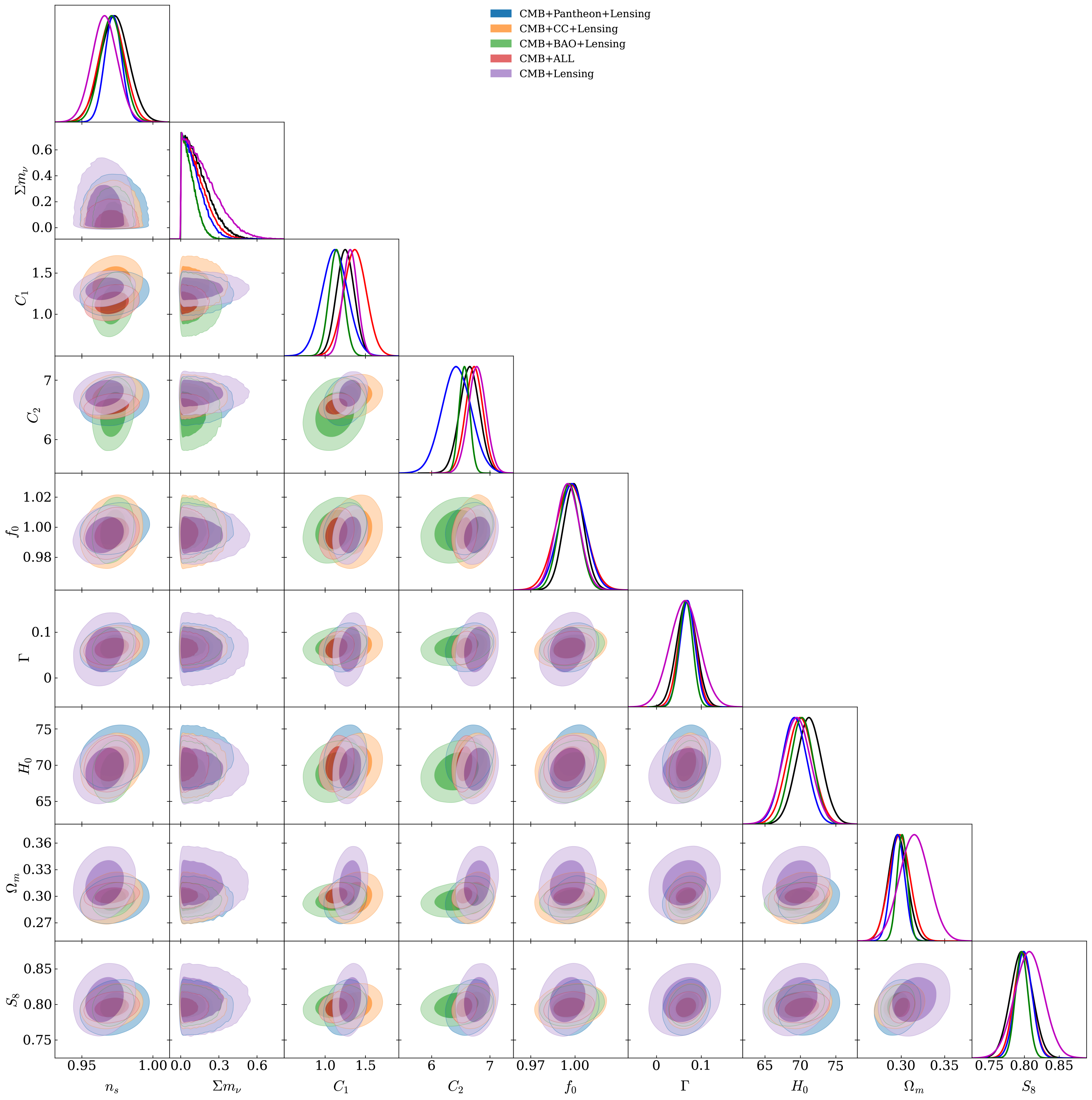}
	\caption{
		Triangle plot showing the marginalized posterior distributions and joint confidence contours for the interacting Hu--Sawicki $f(R)$ gravity model obtained from the combined \textbf{CMB+ALL} dataset analysis. 
	}
	\label{fig:5}
\end{figure}

Finally, introducing a phenomenological energy exchange between the dark sectors within this modified gravity background yields the interacting Hu--Sawicki $f(R)$ scenario (posteriors shown in Fig.~\ref{fig:5}). The baseline and intrinsic modified-gravity variables remain tightly constrained and essentially indistinguishable from the non-interacting case. The distinguishing feature here is the dynamic energy transfer; the joint analysis consistently favors a positive dark-sector coupling, yielding $\Gamma = 0.0655 \pm 0.015$ for the \textbf{CMB+ALL} dataset. This active interaction provides a physical pathway to concurrently ease both the Hubble and $S_8$ tensions. The combination of the modified background and the dark coupling also increases the model's sensitivity to neutrino free-streaming, giving the interacting framework a tighter upper bound on the total neutrino mass, $\Sigma m_\nu < 0.122$~eV.

\begin{table}[H]
	\centering
	\caption{Cosmological parameter constraints for $\Lambda$CDM, Hu--Sawicki $f(R)$, and Interacting Hu--Sawicki $f(R)$+coupling gravity models obtained from different dataset combinations.}
	\label{tab:combined_cosmo_models}
	\resizebox{\textwidth}{!}{
		\begin{tabular}{lc|c|c|c|c}
			\toprule
			\textbf{Parameter} & \textbf{CMB+Pantheon+Lensing} & \textbf{CMB+CC+Lensing} & \textbf{CMB+BAO+Lensing} & \textbf{CMB+ALL} & \textbf{CMB+Lensing} \\
			\midrule
			\hline
			\multicolumn{6}{c}{\textbf{$\Lambda$CDM Model}} \\
			\midrule
			\hline
			$\Omega_b h^2$ & $0.02230 \pm 0.00021$ & $0.02229 \pm 0.00018$ & $0.02234 \pm 0.00020$ & $0.02229 \pm 0.00015$ & $0.02238 \pm 0.00018$ \\
			$\Omega_c h^2$ & $0.1201 \pm 0.0026$ & $0.1198 \pm 0.0022$ & $0.1196 \pm 0.0032$ & $0.1189 \pm 0.0025$ & $0.1202 \pm 0.0033$ \\
			$100\theta_{MC}$ & $1.0409 \pm 0.00062$ & $1.0408 \pm 0.00056$ & $1.0413 \pm 0.0005$ & $1.0409 \pm 0.0003$ & $1.04109 \pm 0.0008$ \\
			$\tau$ & $0.0531 \pm 0.0081$ & $0.0547 \pm 0.009$ & $0.057 \pm 0.006$ & $0.0549 \pm 0.0045$ & $0.055 \pm 0.009$ \\
			$\ln(10^{10} A_s)$ & $3.05 \pm 0.018$ & $3.046 \pm 0.016$ & $3.044 \pm 0.017$ & $3.044 \pm 0.012$ & $3.046 \pm 0.025$ \\
			$n_s$ & $0.966 \pm 0.009$ & $0.969 \pm 0.008$ & $0.973 \pm 0.006$ & $0.968 \pm 0.006$ & $0.966 \pm 0.009$ \\
			$\Omega_m$ & $0.290 \pm 0.013$ & $0.298 \pm 0.011$ & $0.303 \pm 0.009$ & $0.295 \pm 0.007$ & $0.306 \pm 0.017$ \\
			$H_{0}$ & $70.06^{+1.85}_{-1.87}$ & $69.03^{+1.6}_{-1.6}$ & $68.42 \pm 1.42$ & $69.19 \pm 1.42$ & $68.53 \pm 1.8$ \\
			$S_{8}$ & $0.804 \pm 0.016$ & $0.807 \pm 0.014$ & $0.809 \pm 0.013$ & $0.803 \pm 0.009$ & $0.809 \pm 0.018$ \\
			$\chi^{2}$ & $4320.921$ & $2820.325$ & $2804.912$ & $4353.754$ & $2797.286$ \\
			\midrule
			\hline
			\multicolumn{6}{c}{\textbf{Hu--Sawicki $f(R)$ Gravity Model}} \\
			\midrule
			\hline
			$\Omega_b h^2$ & $0.02231 \pm 0.00026$ & $0.02229 \pm 0.00025$ & $0.02234 \pm 0.00018$ & $0.02228 \pm 0.00018$ & $0.02233 \pm 0.00024$ \\
			$\Omega_c h^2$ & $0.1193 \pm 0.0034$ & $0.1191 \pm 0.0040$ & $0.1192 \pm 0.0035$ & $0.1189 \pm 0.0029$ & $0.1195 \pm 0.0038$ \\
			$100\theta_{MC}$ & $1.0409 \pm 0.0007$ & $1.0411 \pm 0.0006$ & $1.0412 \pm 0.0007$ & $1.0409 \pm 0.0006$ & $1.0412 \pm 0.0005$ \\
			$\tau$ & $0.059 \pm 0.008$ & $0.057 \pm 0.006$ & $0.055 \pm 0.006$ & $0.054^{+0.008}_{-0.007}$ & $0.056 \pm 0.008$ \\
			$\ln(10^{10} A_s)$ & $3.049 \pm 0.019$ & $3.048 \pm 0.017$ & $3.045 \pm 0.009$ & $3.048 \pm 0.008$ & $3.051 \pm 0.0219$ \\
			$n_s$ & $0.971 \pm 0.011$ & $0.974 \pm 0.009$ & $0.971 \pm 0.009$ & $0.972 \pm 0.007$ & $0.969 \pm 0.011$ \\
			$C_{1}\,(\times10^{-3})$ & $1.31 \pm 0.12$ & $1.35 \pm 0.16$ & $1.14 \pm 0.15$ & $1.21 \pm 0.10$ & $1.35 \pm 0.17$ \\
			$C_{2}\,(\times10^{-5})$ & $6.43 \pm 0.18$ & $6.52 \pm 0.19$ & $6.39 \pm 0.17$ & $6.51 \pm 0.08$ & $6.68 \pm 0.21$ \\
			$f_0$ & $0.999 \pm 0.009$ & $0.997 \pm 0.012$ & $0.998 \pm 0.009$ & $0.996 \pm 0.008$ & $0.995 \pm 0.011$ \\
			$H_{0}$ & $71.04 \pm 1.82$ & $70.25 \pm 1.71$ & $69.25 \pm 1.42$ & $70.11 \pm 1.32$ & $69.15 \pm 1.91$ \\
			$\Omega_{m}$ & $0.292 \pm 0.012$ & $0.295 \pm 0.010$ & $0.309 \pm 0.010$ & $0.298 \pm 0.008$ & $0.311 \pm 0.013$ \\
			$S_{8}$ & $0.798 \pm 0.014$ & $0.796 \pm 0.012$ & $0.803 \pm 0.011$ & $0.798 \pm 0.008$ & $0.809 \pm 0.022$ \\
			$\chi^{2}$ & $4296.212$ & $2801.432$ & $2794.500$ & $4328.377$ & $2792.258$ \\
			\midrule
			\hline
			\multicolumn{6}{c}{\textbf{Interacting Hu--Sawicki $f(R)$+Coupling Gravity Model}} \\
			\midrule
			\hline
			$\Omega_b h^2$ & $0.02228 \pm 0.00023$ & $0.02230 \pm 0.00022$ & $0.02232 \pm 0.00019$ & $0.02224 \pm 0.00021$ & $0.02232 \pm 0.00021$ \\
			$\Omega_c h^2$ & $0.1196 \pm 0.0038$ & $0.1194 \pm 0.0041$ & $0.1192 \pm 0.0036$ & $0.1190 \pm 0.0027$ & $0.1194 \pm 0.0036$ \\
			$100\theta_{MC}$ & $1.0410 \pm 0.0007$ & $1.0410 \pm 0.0006$ & $1.0413 \pm 0.0007$ & $1.0410 \pm 0.0006$ & $1.04115 \pm 0.0005$ \\
			$\tau$ & $0.061 \pm 0.009$ & $0.059 \pm 0.008$ & $0.057 \pm 0.009$ & $0.056^{+0.007}_{-0.009}$ & $0.056 \pm 0.009$ \\
			$\ln(10^{10} A_s)$ & $3.051 \pm 0.019$ & $3.047 \pm 0.017$ & $3.044 \pm 0.016$ & $3.047^{+0.018}_{-0.019}$ & $3.048 \pm 0.019$ \\
			$n_s$ & $0.973 \pm 0.010$ & $0.971 \pm 0.009$ & $0.972 \pm 0.006$ & $0.971 \pm 0.008$ & $0.966 \pm 0.009$ \\
			$\Sigma m_{\nu}\,(\mathrm{eV})$ & $<0.21$ & $<0.19$ & $<0.18$ & $<0.122$ & $<0.33$ \\
			$C_{1}\,(\times10^{-3})$ & $1.25 \pm 0.11$ & $1.37 \pm 0.14$ & $1.12 \pm 0.16$ & $1.14 \pm 0.09$ & $1.31 \pm 0.09$ \\
			$C_{2}\,(\times10^{-5})$ & $6.65 \pm 0.17$ & $6.73 \pm 0.15$ & $6.026 \pm 0.25$ & $6.56 \pm 0.09$ & $6.78 \pm 0.15$ \\
			$f_0$ & $0.999 \pm 0.007$ & $0.997 \pm 0.010$ & $0.998 \pm 0.009$ & $0.996 \pm 0.007$ & $0.995 \pm 0.008$ \\
			$\Gamma$ & $0.067 \pm 0.022$ & $0.0682 \pm 0.019$ & $0.069 \pm 0.017$ & $0.0655 \pm 0.015$ & $0.063 \pm 0.033$ \\
			$H_{0}$ & $71.21 \pm 1.77$ & $69.95 \pm 1.84$ & $69.16 \pm 1.75$ & $70.21 \pm 1.60$ & $69.45 \pm 1.95$ \\
			$\Omega_{m}$ & $0.296 \pm 0.011$ & $0.298 \pm 0.012$ & $0.296 \pm 0.008$ & $0.301 \pm 0.006$ & $0.315 \pm 0.017$ \\
			$S_{8}$ & $0.796 \pm 0.016$ & $0.798 \pm 0.013$ & $0.799 \pm 0.012$ & $0.796 \pm 0.009$ & $0.807 \pm 0.021$ \\
			$\chi^{2}$ & $4290.824$ & $2796.028$ & $2790.672$ & $4308.312$ & $2788.149$ \\
			\bottomrule
		\end{tabular}
	}
\end{table}

An interesting feature of the interacting-neutrino Hu--Sawicki model is the systematically larger value of the matter density parameter obtained from the \textbf{CMB+Lensing} dataset combination, for which we find $\Omega_m = 0.315 \pm 0.017$, compared with the values preferred by the other dataset combinations. This behavior likely reflects residual parameter degeneracies among $\Omega_m$, the summed neutrino mass $\Sigma m_\nu$, and the modified-gravity parameters. In the absence of complementary geometric probes such as BAO, Pantheon, or CC measurements, the CMB+lensing data permit a substantially broader neutrino-mass range ($\Sigma m_\nu < 0.33~{\rm eV}$) together with a wider region of the modified-gravity parameter space. Since both massive neutrinos and modified gravitational dynamics affect the growth of cosmic structures and the lensing potential, a larger value of $\Omega_m$ may help compensate these effects while maintaining consistency with the observed lensing signal. Once additional low-redshift distance measurements are included, these degeneracies are significantly reduced, leading to tighter constraints and a preferred matter density closer to the canonical value $\Omega_m \simeq 0.30$; this shows that complementary datasets play an important role in breaking parameter degeneracies in interacting modified-gravity cosmologies. The interacting-neutrino model also allows a direct comparison with current cosmological and terrestrial neutrino-mass limits. A widely discussed issue in contemporary neutrino physics is the apparent difference between the stringent upper bounds on the summed neutrino mass, $\Sigma m_\nu$, inferred from cosmological observations and the comparatively weaker constraints obtained from direct laboratory measurements. Cosmological analyses based on CMB and large-scale structure data typically favor sub-eV upper limits on $\Sigma m_\nu$ \cite{Planck2018,Jiang2024}, whereas direct kinematic experiments, such as KATRIN, currently constrain the effective electron-neutrino mass to $m_\beta < 0.45\,\mathrm{eV}$ (95\% C.L.) \cite{Lokhov2022,Aker2024}. Our results show that the interacting-neutrino scenario remains fully compatible with both cosmological and laboratory constraints. Depending on the adopted dataset combination, we obtain upper limits in the range $\Sigma m_\nu < 0.122$--$0.33\,\mathrm{eV}$, with the most stringent bound arising from the CMB+ALL dataset. These limits are comparable to, and in several cases more restrictive than, current direct laboratory constraints, while remaining consistent with all available observations. The presence of the neutrino--gravity interaction enlarges the allowed neutrino-mass parameter space relative to the standard cosmological scenario and permits larger values of $\Sigma m_\nu$ without a significant deterioration in the goodness of fit. Although the resulting bounds do not completely resolve the difference between cosmological and terrestrial neutrino-mass determinations, they suggest that neutrino interactions can partially alleviate this discrepancy and provide a physically motivated extension of the standard cosmological framework. This behavior is broadly consistent with previous studies of interacting neutrino cosmologies \cite{Lesgourgues2014}.

\subsection{Quantification and Mitigation of $H_0$ and $S_8$ Tensions}
\label{subsec:tensions_mitigation}

A cosmological model should fit individual datasets and also reconcile the persistent discrepancies between early- and late-Universe observations. To evaluate this, we first examine the internal consistency of the models by comparing the constraints derived from the late-time dataset (DC1: CC+SN) against those from the early-Universe dataset (DC2: BAO+CMB+Lensing).

Table~\ref{tab:h0_internal_tension} summarizes the internal compatibility regarding the Hubble constant. The well-known tendency of early-Universe probes to favor lower expansion rates is evident across all frameworks. However, the statistical discrepancy between DC1 and DC2 decreases significantly when moving beyond standard General Relativity. Specifically, the internal $H_0$ tension drops from $1.27\sigma$ in the baseline $\Lambda$CDM model to $0.82\sigma$ in the non-interacting Hu--Sawicki framework, and further improves to $0.68\sigma$ when the dark-sector interaction is introduced. This trend indicates that the geometric modifications and the energy exchange mechanism provide a physical channel that improves compatibility between low- and high-redshift dynamics.
\begin{table}[H]
	\centering
	\caption{Comparison of the inferred Hubble constant values obtained from the DC1 (CC+SN) and DC2 (BAO+CMB+Lensing) dataset combinations for the three cosmological models considered in this work. The last column reports the statistical compatibility between the two dataset combinations, computed as
		\(
		T=\frac{|H_0^{\rm DC1}-H_0^{\rm DC2}|}
		{\sqrt{\sigma_{\rm DC1}^2+\sigma_{\rm DC2}^2}}.
		\)
	}
	\label{tab:h0_internal_tension}
	\resizebox{\textwidth}{!}{
		\begin{tabular}{|c|c|c|c|}
			\hline
			\textbf{Model} &
			\textbf{$H_0(\mathrm{km\,s^{-1}Mpc^{-1}})$ DC1} &
			\textbf{$H_0(\mathrm{km\,s^{-1}Mpc^{-1}})$ DC2} &
			\textbf{DC1--DC2 Compatibility} \\
			\hline
			
			Flat $\Lambda$CDM
			&
			$71.21\pm1.68$
			&
			$68.42\pm1.42$
			&
			$1.27\sigma$
			\\
			\hline
			
			Hu--Sawicki $f(R)$ (non-interacting)
			&
			$71.14\pm1.81$
			&
			$69.25\pm1.42$
			&
			$0.82\sigma$
			\\
			\hline
			
			Interacting Hu--Sawicki $f(R)$
			&
			$70.92\pm1.92$
			&
			$69.16\pm1.75$
			&
			$0.68\sigma$
			\\
			\hline
			
	\end{tabular}}
\end{table}

Similarly, we assess the clustering amplitude $S_8$ across these individual dataset combinations in Table~\ref{tab:s8_tension_compare1}. Unlike $H_0$, the inferred $S_8$ values remain relatively stable when switching from DC1 to DC2. For instance, in the interacting Hu--Sawicki scenario, $S_8$ marginally shifts from $0.797 \pm 0.018$ (DC1) to $0.799 \pm 0.012$ (DC2). Despite this stability, early-Universe probes (DC2) generally yield slightly higher tensions with weak-lensing measurements (KiDS and DES-Y6) compared to late-time probes (DC1). Nevertheless, across both individual data splits, the modified gravity models systematically prefer lower $S_8$ values than $\Lambda$CDM, hinting at a generalized suppression of structure growth.
\begin{table}[H]
	\centering
	\caption{Comparison of the inferred $S_8$ values and their tensions with \textit{Planck} 2018, KiDS, and DES-Y6 measurements for all cosmological models considered in this work.}
	\label{tab:s8_tension_compare1}
	\resizebox{\textwidth}{!}{
		\begin{tabular}{|c|c|c|c|c|c|}
			\hline
			\textbf{Model} & \textbf{Dataset} & \textbf{$S_8$} & \textbf{Tension with Planck 2018} & \textbf{Tension with KiDS} & \textbf{Tension with DES-Y6} \\
			\hline
			Flat $\Lambda$CDM & DC1  & $0.801 \pm 0.019$ & $1.18\sigma$ & $1.45\sigma$ & $0.97\sigma$ \\
			\hline
			Flat $\Lambda$CDM & DC2  & $0.809 \pm 0.013$ & $1.26\sigma$ & $1.96\sigma$ & $1.41\sigma$ \\
			\hline
			Hu--Sawicki $f(R)$ (non-interacting) & DC1  & $0.798 \pm 0.017$ & $1.49\sigma$ & $1.52\sigma$ & $0.90\sigma$ \\
			\hline
			Hu--Sawicki $f(R)$ (non-interacting) & DC2  & $0.803 \pm 0.011$ & $1.61\sigma$ & $2.00\sigma$ & $1.40\sigma$ \\
			\hline
			Interacting Hu--Sawicki $f(R)$ & DC1  & $0.797 \pm 0.018$ & $1.44\sigma$ & $1.43\sigma$ & $0.92\sigma$ \\
			\hline
			Interacting Hu--Sawicki $f(R)$ & DC2  & $0.799 \pm 0.012$ & $1.76\sigma$ & $1.83\sigma$ & $1.12\sigma$ \\
			\hline
	\end{tabular}}
\end{table}

Beyond internal consistency, we also quantify the residual tensions with independent external calibrations using the complete joint dataset (CMB+ALL). Table~\ref{tab:H0_tension} reports the tension metrics for $H_0$. While the standard $\Lambda$CDM model remains strongly anchored to the \textit{Planck} early-Universe calibration---leaving a residual $2.3\sigma$ tension with the local SH0ES (R22) measurement---both modified-gravity extensions systematically drive $H_0$ toward higher values. The interacting Hu--Sawicki scenario shows the largest reduction, lowering the $H_0$ tension with R22 to $1.4\sigma$.

\begin{table}[H]
	\centering
	\caption{Constraints on $H_0$ and corresponding tension estimates with the \textit{Planck} 2018 and R22 measurements for different cosmological models and dataset combinations.}
	\label{tab:H0_tension}
	\resizebox{\textwidth}{!}{
		\begin{tabular}{|c|c|c|c|c|}
			\hline
			Model & Dataset Combination & $H_0~[{\rm km~s^{-1}Mpc^{-1}}]$ & $T(H_0)_{\rm Planck}$ & $T(H_0)_{\rm R22}$ \\
			\hline
			$\Lambda$CDM & Full Dataset & $69.19\pm1.42$ & $1.1\sigma$ & $2.3\sigma$ \\
			\hline
			Hu--Sawicki $f(R)$ & Full Dataset & $70.11\pm1.32$ & $1.9\sigma$ & $1.5\sigma$ \\
			\hline
			Interacting Hu--Sawicki $f(R)$ & Full Dataset & $70.21\pm1.60$ & $1.7\sigma$ & $1.4\sigma$ \\
			\hline
	\end{tabular}}
\end{table}

Concurrently, the inclusion of modified gravitational dynamics systematically favors a lower clustering amplitude for the full dataset, as detailed in Table~\ref{tab:S8_tension}. Relative to the $\Lambda$CDM baseline ($S_8 = 0.803 \pm 0.009$), the interacting framework yields $S_8 = 0.796 \pm 0.009$. This moderation reduces the residual $S_8$ tension with DES-Y6 below the $1\sigma$ threshold ($0.9\sigma$) and lowers the discrepancy with KiDS to $1.2\sigma$, while remaining statistically consistent with the \textit{Planck} 2018 primary CMB calibration (within $1.9\sigma$).

\begin{table}[H]
	\centering
	\caption{Constraints on $S_8$ and corresponding tension estimates with the \textit{Planck} 2018, DES-Y6, and KiDS  measurements for different cosmological models.}
	\label{tab:S8_tension}
	\resizebox{\textwidth}{!}{
		\begin{tabular}{|c|c|c|c|c|c|}
			\hline
			Model & Dataset Combination & $S_8$ & $T(S_8)_{\rm Planck}$ & $T(S_8)_{\rm DES}$ & $T(S_8)_{\rm KiDS}$ \\
			\hline
			$\Lambda$CDM & Full Dataset & $0.803\pm0.009$ & $1.8\sigma$ & $1.3\sigma$ & $1.6\sigma$ \\
			\hline
			Hu--Sawicki $f(R)$ & Full Dataset & $0.798\pm0.008$ & $2.0\sigma$ & $1.0\sigma$ & $1.3\sigma$ \\
			\hline
			Interacting Hu--Sawicki $f(R)$ & Full Dataset & $0.796\pm0.009$ & $1.9\sigma$ & $0.9\sigma$ & $1.2\sigma$ \\
			\hline
	\end{tabular}}
\end{table}

These results show that combining late-time geometric modifications with an effective dark-sector energy exchange provides a phenomenological pathway to a simultaneous, albeit moderate, alleviation of both the cosmic expansion ($H_0$) and clustering ($S_8$) tensions, without compromising the fit to early-Universe observations.

\subsection{Model comparison using the full dataset combination}

The statistical comparison of the cosmological models using the full dataset combination ($\mathrm{CC+SN+BAO+CMB+Lensing}$) is summarized in Table~\ref{tab:AIC_BIC_models0}. In addition to the minimum chi-square values, the Akaike and Bayesian information criteria (AIC and BIC) are reported as complementary model-selection diagnostics that account for the balance between goodness of fit and model complexity (see, e.g., Refs.~\cite{akaike1974,schwarz1978}).

\begin{table}[H]
	\centering
	\caption{Statistical comparison between the cosmological models using the full dataset combination ($\mathrm{CC+SN+BAO+CMB+Lensing}$). The quoted $\Delta$AIC and $\Delta$BIC values are computed relative to the model with the minimum information-criterion value.}
	\label{tab:AIC_BIC_models0}
	\resizebox{0.9\textwidth}{!}{
		\begin{tabular}{|c|c|c|c|c|c|c|}
			\hline
			Model & Free Parameters ($k$) & $\chi^2_{\rm min}$ & AIC & BIC & $\Delta{\rm AIC}$ & $\Delta{\rm BIC}$ \\
			\hline
			$\Lambda$CDM
			& 3
			& 4353.754
			& 4359.754
			& 4378.21
			& 35.44
			& 4.68 \\
			\hline
			Hu--Sawicki $f(R)$
			& 6
			& 4328.377
			& 4340.377
			& 4377.29
			& 16.07
			& 3.76 \\
			\hline
			Interacting Hu--Sawicki $f(R)$
			& 8
			& 4308.312
			& 4324.312
			& 4373.53
			& 0.00
			& 0.00 \\
			\hline
		\end{tabular}
	}
\end{table}

Both Hu--Sawicki $f(R)$ scenarios improve on the $\chi^2_{\rm min}$ of the standard $\Lambda$CDM model, with the interacting framework giving the lowest value overall. The AIC favors the interacting model, and although the stronger complexity penalty of the BIC narrows the gap between the three scenarios, the ranking is unchanged: the interacting Hu--Sawicki model remains the preferred description of the full dataset combination, consistent with the parameter constraints reported in Sec.~V.

\section{Conclusion}

In this work, we investigated viable extensions of the Hu--Sawicki \(f(R)\) gravity framework, focusing on the combined effects of massive neutrinos and an effective dark-sector interaction. The analysis was motivated by the persistent discrepancies between early- and late-Universe measurements, particularly the \(H_0\) and \(S_8\) tensions, and aimed to assess whether these additional physical ingredients can improve the agreement between cosmological observations while remaining consistent with current data. Starting from the metric formulation of \(f(R)\) gravity, we derived the corresponding background and perturbation equations and implemented the resulting cosmological models within the \texttt{MontePython--hi\_class} framework. Cosmological constraints were obtained through Bayesian Markov Chain Monte Carlo analyses using complementary observational probes, including Planck 2018 CMB data, CMB lensing, BAO measurements, Cosmic Chronometers, and the Pantheon supernova sample. The observational results indicate that the Hu--Sawicki extensions remain compatible with all dataset combinations considered in this work. Relative to the standard \(\Lambda\)CDM model, the interacting scenario shifts the preferred value of the Hubble constant toward larger values and yields slightly lower values of the clustering parameter \(S_8\) across all dataset combinations. For the full dataset combination, we obtain $H_0 = 70.21 \pm 1.60~{\rm km~s^{-1}Mpc^{-1}}$ and $S_8 = 0.796 \pm 0.009,$ compared with $H_0 = 69.19 \pm 1.42~{\rm km~s^{-1}Mpc^{-1}}$ and $S_8 = 0.803 \pm 0.009$ in the \(\Lambda\)CDM framework. These shifts reduce, although do not completely remove, the residual discrepancies with local measurements of the expansion rate and recent weak-lensing observations. From a statistical perspective, the interacting Hu--Sawicki model provides the best overall fit among the scenarios examined. For the full dataset combination, the model achieves an improvement of approximately $\Delta\chi^2 \simeq -45$ relative to \(\Lambda\)CDM and yields the minimum AIC and BIC values reported in Table~\ref{tab:AIC_BIC_models0}, indicating that the gain in likelihood outweighs the added complexity of the interaction and neutrino sectors. The inferred Hu--Sawicki parameters remain close to the General Relativity limit, implying that current observations allow only moderate departures from standard gravity, while still leaving the model enough flexibility to modify late-time expansion and the growth of matter perturbations. The neutrino sector is also consistent with current cosmological constraints, with the strongest bound obtained from the full dataset combination, $\sum m_\nu < 0.122~{\rm eV},$ compatible with current cosmological and laboratory limits. The interaction parameter remains positive across all dataset combinations, $\Gamma \approx 0.063$--$0.069$; although our treatment is phenomenological, this may point to an effective energy exchange within the dark sector, with consequences for both the background expansion and the growth of structure. Taken together, these results make the interacting Hu--Sawicki model a statistically competitive alternative to the standard cosmological model, capable of simultaneously shifting \(H_0\) upward and mildly suppressing \(S_8\) while remaining consistent with current observations. The present results do not amount to definitive evidence for modified gravity or dark-sector interactions, but they show such extensions remain viable and worth pursuing with higher-precision data. Future work could test these predictions against nonlinear structure formation and redshift-space distortions using forthcoming DESI, \textit{Euclid}, and Rubin LSST data.

\appendix

\section{Dynamical System of Equations}

Since Eqs.~(15--17) constitute a coupled nonlinear system that generally does not admit closed-form analytical solutions, we reformulate them as an autonomous system of first-order differential equations by introducing suitable dimensionless variables. This phase-space representation provides a powerful analytical framework for investigating the stability, asymptotic behavior, and critical points of the interacting $f(R)$ cosmology. 

It is important to emphasize that the cosmological parameter estimation carried out in Section IV using \texttt{MontePython} and \texttt{hi\_class} is based directly on the exact background and linear perturbation equations derived from the modified gravitational action, rather than this dimensionless system. Therefore, the autonomous formulation presented here is introduced primarily as a complementary theoretical tool. It provides physical intuition regarding the underlying dynamics---such as identifying equilibrium points that correspond to distinct cosmological epochs---and theoretically guarantees that the modified parameter space possesses a stable late-time accelerating attractor.

The new variables introduced for this transformation are generally defined as follows:

\begin{align}\label{eq4}
	&\eta_{1}=\frac{\phi^{\prime}}{\phi H}, \eta_{2}=\frac{\kappa}{H}, \eta_{3}=\frac{f^{\prime}_{R}}{H(1+f_{R})}, \eta_{4}=\frac{\delta}{\phi},\nonumber \\&
	\eta_{5}=\frac{\rho_{m}a^{2}}{(1+f_{R})}, \eta_{6}=\frac{\Psi^{\prime}}{\phi H}, \eta_{7}=\frac{\Psi}{\phi}, \eta_{8}=\frac{\rho_{\nu}a^{2}}{(1+f_{R})},
\end{align}

Using the Hu-Sawicki form of \( f(R) \), the previously defined autonomous equations are modified as follows. These equations describe the evolution of key dynamical variables in the system:

\begin{itemize}
	\item \textbf{Evolution of \( \eta_1 \) (Scalar Field Normalized Derivative):}
	\begin{equation}
		\frac{d\eta_1}{dN} = \chi_3 - \eta_1^2 - \eta_1
	\end{equation}
	where \( \eta_1 \) represents the normalized derivative of the scalar field.
	
	\item \textbf{Evolution of \( \eta_2 \) (Curvature-Hubble Ratio):}
	\begin{equation}
		\frac{d\eta_2}{dN} = -\eta_2 \chi_1
	\end{equation}
	where \( \eta_2 \) quantifies the ratio of the Ricci curvature to the Hubble parameter.
	
	\item \textbf{Evolution of \( \eta_3 \) (Modified Gravity Parameter):}
	\begin{equation}
		\frac{d\eta_3}{dN} = \beta - \eta_3^2 - \chi_1 \eta_3
	\end{equation}
	where \( \eta_3 \) is defined as:
	\begin{equation}
		\eta_3 = \frac{f'_R}{H(1+f_R)}
	\end{equation}
	and the function \( \beta \) is given by:
	\begin{equation}
		\beta = \frac{f''_R}{(1+f_R)H} - 1
	\end{equation}
	with \( f_R \) and \( f''_R \) derived from the Hu-Sawicki form of \( f(R) \).
	The parameter ($\beta$) plays a key role in characterizing deviations from General Relativity within ($f(R)$) gravity. Through its dependence on ($f_{RR}\equiv d^2f/dR^2$), it encodes the dynamics of the additional scalar degree of freedom introduced by the modified gravitational action. As such, ($\beta$) provides a convenient measure of the relative importance of the modified-gravity sector compared with the standard cosmological expansion. In the limit ($\beta=-1$), the modifications vanish and General Relativity is effectively recovered. The evolution of ($\beta$) therefore directly influences both the background expansion history and the growth of cosmological perturbations, making it a useful quantity for assessing the observational signatures of ($f(R)$) models.
	\item \textbf{Evolution of \( \eta_4 \) (Scalar Field Perturbation):}
	\begin{equation}
		\frac{d\eta_4}{dN} = \chi_4\eta_{4} (1 - \eta_1)
	\end{equation}
	where \( \eta_4 \) governs perturbations in the scalar field.
	
	\item \textbf{Evolution of \( \eta_5 \) (Matter Density Contribution):}
	\begin{equation}
		\frac{d\eta_5}{dN} = -\eta_5 (1 + \eta_3)
	\end{equation}
	where \( \eta_5 \) describes the evolution of the matter density.
	
	\item \textbf{Evolution of \( \eta_6 \) (Gravitational Potential Evolution):}
	\begin{equation}
		\frac{d\eta_6}{dN} = \chi_2 - \eta_6 (\eta_1 + \chi_1)
	\end{equation}
	where \( \eta_6 \) characterizes the evolution of the gravitational potential.
	
	\item \textbf{Evolution of \( \eta_7 \) (Additional Gravitational Potential Parameter):}
	\begin{equation}
		\frac{d\eta_7}{dN} = \eta_6 - \eta_7 \eta_1
	\end{equation}
	where \( \eta_7 \) is another parameter related to the gravitational potential.
	
	\item \textbf{Evolution of \( \eta_8 \) (Neutrino Contribution):}
	\begin{equation}
		\frac{d\eta_8}{dN} = \eta_8 (3 \omega_\nu - 1) - \eta_3 \eta_8 + \Gamma \eta_2 \eta_8
	\end{equation}
	where \( \eta_8 \) incorporates neutrino effects, and \( \Gamma \) represents a coupling term between neutrinos and the gravitational sector.
\end{itemize}
The behavior of neutrinos—elusive and enigmatic particles—is significantly influenced by modifications to gravitational dynamics, such as those introduced in perturbed \( f(R) \) gravity \cite{Nojiri2011}. In this section, we analyze the constraints on the total mass of neutrinos within this framework. A key equation governing the energy density of neutrinos (\( \rho_{\nu} \)), the scale factor (\( a \)), and the modification term (\( 1+f_{R} \)) is given by \cite{Sotiriou2010}:

\begin{equation}
	\eta_{8} = \frac{\rho_{\nu} a^{2}}{(1+f_{R})}.
\end{equation}

To extract meaningful constraints on the total neutrino mass, additional physical assumptions and observational inputs are required. The energy density of neutrinos can be expressed in terms of their mass (\( m_{\nu} \)) and temperature (\( T_{\nu} \)). A common approach utilizes the Fermi-Dirac distribution for relativistic neutrinos, yielding \cite{Kolb1990,Lesgourgues2006}:

\begin{equation}
	\rho_{\nu} = \frac{7\pi^2}{120} g_{\nu} T_{\nu}^4,
\end{equation}

where \( g_{\nu} \) denotes the number of degrees of freedom for neutrinos, typically taken as 2 per neutrino species. Substituting this expression into Equation (49), we obtain:

\begin{equation}
	\eta_{8} = \frac{\left(\frac{7\pi^2}{120} g_{\nu} T_{\nu}^4\right) a^{2}}{(1+f_{R})}.
\end{equation}

The sum of neutrino masses, a fundamental parameter in cosmology, is linked to the neutrino density parameter \( \Omega_{\nu} \) via the following relationship \cite{Lesgourgues2006}:

\begin{equation}
	\Omega_{\nu} = \frac{\sum m_{\nu}}{94 h^2}.
\end{equation}

To determine the total mass of neutrinos, \( \sum m_{\nu} \), we must accurately estimate the cosmological parameters \( \Omega_{\nu} \) and \( h \). This is typically achieved through Bayesian inference techniques, specifically using the Markov Chain Monte Carlo (MCMC) method.


To facilitate the dynamical-system analysis of Eqs.~(A2), (A3),(A4), (A7)--(A11), we define the following auxiliary variables:

\begin{equation}\label{is}
	\chi_{1}=\frac{\mathcal{H}^{\prime}}{\mathcal{H}^{2}}, \qquad
	\chi_{2}=\frac{\Psi^{\prime\prime}}{\phi H^{2}}, \qquad
	\chi_{3}=\frac{\phi^{\prime\prime}}{\phi H^{2}}, \qquad
	\chi_{4}=\frac{\delta^{\prime}}{\phi H}.
\end{equation}

The evolution equations for these variables are then obtained by differentiating the above definitions and combining them with the perturbation equations, leading to a closed dynamical system.
\color{black}
Now, for the autonomous equations of motions, we obtain

\begin{equation}\label{is}
	\begin{split}
		\chi_{1}=\frac{1}{1-\eta_{7}}[\eta_{1}+\eta_{6}+\frac{1}{3}\eta_{2}^{2}(1+\eta_{7})+(3-\eta_{7}+\eta_{6})\eta_{5}+\\(3-\eta_{7}+\eta_{6})\eta_{8}-\frac{\kappa^{2}}{k^{2}}\eta_{5}\eta_{4}\eta_{2}^{2}]
	\end{split}
\end{equation}

\begin{equation}\label{eq4}
	\begin{split}
		\chi_{2}=\frac{-2}{1-\eta_{7}}[\eta_{1}+\eta_{6}+\frac{1}{3}\eta_{2}^{2}(1+\eta_{7})+\\(3-\eta_{7}+\eta_{6})\eta_{5}- \frac{\kappa^{2}}{k^{2}}\eta_{5}\eta_{4}\eta_{2}^{2}] \\
		-\eta_{1}-3\eta_{6}+\frac{1}{3}\eta_{2}^{2}-\eta_{6}\eta_{1}+\\ \frac{1}{3}\eta_{2}^{2}(1-2\eta_{7})+\frac{\Omega}{3}(1-\eta_{7})\frac{1}{k^{2}}\eta_{2}^{2}
	\end{split}
\end{equation}

\begin{equation}\label{is}
	\begin{split}
		\chi_{3}=-\chi_{2}-3\chi_{1}(1+\frac{1}{3}\eta_{7})-3\eta_{1}-3\eta_{6}-2\eta_{7}-\\ (3-\eta_{7}+3\eta_{1})\eta_{5}+\beta(\eta_{7}-3)
	\end{split}
\end{equation}

\begin{equation}\label{is}
	\chi_{4}=\frac{-\kappa^{2}[(2-\eta_{7})\eta_{3}+\eta_{1}+\eta_{6}+1+\eta_{7}]}{\kappa^{2}\eta_{5}}+3\eta_{6}
\end{equation}
The parameter \( \chi_1 \) plays a crucial role in cosmological analysis, as it facilitates the expression of fundamental cosmological parameters, such as the deceleration parameter \( q \) and the effective equation of state parameter \( w_{\text{eff}} \), in terms of its contributions. These parameters provide essential insights into the dynamical evolution of the Universe, particularly in understanding cosmic acceleration and the nature of dark energy.  

The deceleration parameter (q) is a dimensionless quantity that characterizes the expansion dynamics of the Universe. In terms of the conformal Hubble parameter ($\mathcal H$), it is defined as
\begin{equation}
	q=-1-\frac{\mathcal H'}{\mathcal H^2},
\end{equation}
where a prime denotes differentiation with respect to conformal time. Negative values of (q) correspond to an accelerating Universe, whereas positive values indicate a decelerating expansion.

The effective equation-of-state parameter describing the total cosmic fluid is given by
\begin{equation}
	w_{\rm eff}=-1-\frac{2}{3}\frac{\mathcal H'}{\mathcal H^2}.
\end{equation}
This quantity provides a convenient measure of the dominant energy component at a given cosmological epoch.

For completeness, the deceleration parameter can equivalently be written in terms of the scale factor (a(t)) as
\begin{equation}
	q=-\frac{a\ddot a}{\dot a^{,2}},
\end{equation}
where an overdot denotes differentiation with respect to cosmic time (t).

\end{document}